\documentclass[9pt,shortpaper,twoside,web]{ieeecolor}
\usepackage{generic}
\usepackage{cite}
\usepackage{amsmath,amssymb,amsfonts}
\usepackage{algorithmic}
\usepackage{graphicx}
\usepackage{textcomp}
\usepackage{epstopdf}
\usepackage{subfigure}
\def\BibTeX{{\rm B\kern-.05em{\sc i\kern-.025em b}\kern-.08em
    T\kern-.1667em\lower.7ex\hbox{E}\kern-.125emX}}
\begin{document}
\title{Distributed Event-triggered Control of Networked Strict-feedback Systems Via Intermittent State Feedback}
\author{Libei~Sun,~Xiucai~Huang,~Yongduan~Song,~\IEEEmembership{Fellow,~IEEE}}
\maketitle

\begin{abstract}
It poses technical difficulty to achieve stable tracking even for single mismatched nonlinear strict-feedback systems when intermittent state feedback is utilized. The underlying problem becomes even more complicated if such systems are networked with directed communication and state-triggering setting. In this work, we present a fully distributed neuroadaptive tracking control scheme for multiple agent systems in strict-feedback form using triggered state from the agent itself and the triggered states from the neighbor agents. To circumvent the non-differentiability of virtual controllers stemming from state-triggering, we first develop a distributed continuous control scheme under regular state feedback, upon which we construct the distributed event-triggered control scheme by replacing the states in the preceding scheme with the triggered ones. Several useful lemmas are introduced to allow the stability condition to be established with such replacement, ensuring that all the closed-loop signals are semi-globally uniformly ultimately bounded (SGUUB), with the output tracking error converging to a residual set around zero. Besides, with proper choices of the design parameters, the tracking performance in the mean square sense can be improved. Numerical simulation verifies the benefits and efficiency of the proposed method.
\end{abstract}
%\IEEEpeerreviewmaketitle

\section{Introduction}
%\IEEEPARstart
\allowdisplaybreaks
Nowadays control systems are normally implemented over networks \cite{gupta2008overview}.
As such, it is highly desirable to share network resources in both sensor/actuator and control input channels to save on board communication bandwidth and stored energy, both are limited in networked systems operating autonomously. Under such setting, preserving stability under certain communication and energy constraints is vital in networked control systems. A commonly used data transmission/communication approach in traditional digital control techniques is the fixed time-scheduled sampling, which, however,
%cause unnecessary overloads the communication networks \cite{aastrom1999comparison}.
causes unnecessary overloads to the communication networks \cite{aastrom1999comparison}.

The emergence of event-triggered control has provided an appealing avenue for effective transmissions of measured signals through certain triggering conditions \cite{astrom2008event}. Early results of event-triggered control mainly focus on linear systems, see \cite{SEURET201647,heemels2012periodic} and the references therein. The work in \cite{tabuada2007event} further considers a class of nonlinear systems in the framework of event-triggered control, while the input-to-state stability (ISS) condition is required. Such limitation is successfully removed in \cite{xing2016event,zhang2021adaptive} by developing some event-triggered adaptive control algorithms based on the backstepping technique. However, the results in \cite{SEURET201647,heemels2012periodic,tabuada2007event,xing2016event,zhang2021adaptive} are only applicable for single nonlinear systems, albeit there exist a large of engineering systems that are literally networked.

On the other hand, most networked multiple agent systems are usually in short of communication and energy resources, especially when the agent itself or its internal devices are battery-powered, wherein the communication bandwidth and channels among the connected subsystems are limited, motivating extensive and in-depth studies on distributed and event-triggered control
methodologies \cite{6068223,wang2020adaptive}, which can be broadly classified two categories: one is referred to as input-triggered control, and the other as state-triggered control. In input-triggered control, the frequency of updating the actuation signal is substantially reduced, see \cite{8474310,cao2019event} and the references therein, yet the communication resources wasted among agents are ignored intentionally. As for distributed state-triggered control, some efforts have been made for linear systems \cite{ding2016distributed}, first-order nonlinear systems \cite{ZHAN2019104531}, and second-order nonlinear systems \cite{7006773}. Some other works have also considered high-order nonlinear systems \cite{8319522,8910377}, those methods, however, rely on a key requirement that the neighbors' states are continuously monitored to implement the designed triggering condition.
Towards relaxing such a restriction, an effort is made in \cite{seyboth2013event} through a state-triggered control strategy for integrator network systems, while the communication graph is required to be undirected. For the directed topology, a distributed state-triggered adaptive controller is designed in \cite{wang2021adaptive} for a class of norm-form nonlinear systems. The problem of state-triggered output consensus tracking is addressed by using backstepping based output feedback in \cite{long2022output}. Nevertheless, in most of the above results, the explosion of complexity issue always exists owing to repetitive/recursive differentiation of virtual controllers in the backstepping technique.
Moreover, in the framework of event-triggered control, results on networked nonlinear strict-feedback systems with mismatched uncertainties are still limited and the related issues have not been well addressed.

Motivated by the above analysis and discussion, in this work we develop a fully distributed neuroadaptive control scheme for a class of networked mismatched nonlinear strict-feedback systems with directed communication topology and state-triggering setting. The key contributions and features of our method can be summarized as follows:
\begin{itemize}
\item [i)]{By considering non-identical agent model with mismatched uncertainties, the developed control scheme is applicable to a larger class of multiple agent systems, covering those studies on \cite{wang2020adaptive}, \cite{ZHAN2019104531,7006773}, \cite{seyboth2013event,wang2021adaptive,long2022output},  wherein the models are in low-order forms or in normal forms with parametric uncertainties. To our best knowledge, this is the first solution to the distributed adaptive control problem for networked strict-feedback multiple agent systems with state-triggering setting.}
\item [ii)]{The proposed approach is more effective in saving communication and energy resources as compared with the existing ones \cite{wang2020adaptive}, \cite{wang2021adaptive,long2022output}, because the designed triggering mechanism allows all the states sensoring and data transmission to be executed intermittently on the event-driven basis, and the states include those from the agent itself, the leader, and the neighbors (not just the states between subsystems), thus the sensors do not need to be powered all the time and the data from the sensors to the controllers does not have to be transmitted ceaselessly.}
\item [iii)]{As the intermittent states make it invalid to take the differentiation of virtual controller as required in backstepping design, we choose to develop a fully distributed continuous control scheme first using regular state feedback, upon which we construct the distributed event-triggered control scheme by replacing the states in the preceding scheme with the triggered ones, which is still fully  distributed since only local information from the neighbors is involved. Several lemmas are established to authenticate that such replacement preserves consensus tracking stability, while at the same time significantly saving communication and energy resources.}
\end{itemize}

The remainder of the paper is organized as follows. In Section II, the problem is formally stated and the preliminaries are given. In Section III, a distributed continuous control scheme is firstly designed, upon which the distributed event-triggered control scheme is then proposed via intermittent state feedback in Section IV. Section V presents the simulation results. Section VI concludes the paper.

\section{Preliminaries and Problem Formulation}
\subsection{Basic Graph Theory}
A directed graph among $N$ agents is denoted by $\mathcal{G}=\{\mathcal{V}, \mathcal{E}, \mathcal{A}\}$, which consists of a set of nodes $\mathcal{V}= \{v_1,\cdots,v_ N\}$, a group of edges $\mathcal{E}$ and a weighted adjacency matrix ${\mathcal{A}}=[a_{ij}]\in \mathcal{R}^{N \times N}$. If there is an edge $(i,j)$ between nodes $i$ and $j$, then $i$ and $j$ are called adjacent, i.e., $\mathcal{E} =\{(i,j)\in \mathcal{V}\times \mathcal{V}:i,j \,\,{\rm{adjacent}}\}$.
An edge $(v_i,v_j)\in \mathcal{E}$ suggests that node $j$ can receive information from node $i$. For adjacency matrix $\mathcal{A}$, if $(v_i,v_j)\in \mathcal{E}$, then $a_{ij}>0$, otherwise, $a_{ij}=0$. $\mathcal{B}= {\rm{diag}}\{\mu_1, \cdots, \mu_n\}$ is the leader adjacency matrix, where $\mu_1>0$ if there is a directed edge from the leader to the agent i, and $\mu_1=0$ otherwise. It is assumed that there are no self-loops or parallel edges in the network. If there is a path from $i$ to $j$, then $i$ and $j$ are called connected. If all pairs of nodes in $\mathcal{G}$ are connected, then $\mathcal{G}$ is called connected.  $\mathcal{N}_i=\{v_j\in \mathcal{V}|{\left(v_j,v_i\right)}\in \mathcal{E},i \ne j\}$ is the indices set of neighbors of node $v_i$. The Laplacian matrix $\mathcal{L}=[l_{ij}]\in \mathcal{R}^{N\times N}$ is defined as $l_{ii}=\sum_{j=1, j \neq i}^{N} a_{i j}$ and $l_{i j}=-a_{i j}, i \neq j$, satisfying $\mathcal{L}=\mathcal{D}-\mathcal{A}$, where $\mathcal{D}={\rm{diag}}(d_1,\cdots,d_N)\in \mathcal{R}^{N\times N}$ is the absolute in-degree matrix with $d_{i}=\sum_{j\in N_i} a_{ij}$.

\subsection{Problem Formulation}
Consider a multiple agent system composed of $N\,(N\ge{2})$ agents under a directed topology $\mathcal{G}$, with the $i$th, $i=1,\cdots,N$ agent modeled as:
\begin{flalign}
{{\dot x}_{i,k}} =\,& {x_{i,k + 1}} + {f_{i,k}}\left( {{x_{i,1}}, \cdots ,{x_{i,k}}} \right),\,k = 1,2, \cdots ,n - 1 &  \nonumber \\
{{\dot x}_{i,n}} =\,& {u_i} + {f_{i,n}}\left( {{x_{i,1}}, \cdots ,{x_{i,n}}}\right)&\nonumber \\
{y_i} =\,& {x_{i,1}} & \label{eq:1}
\end{flalign}
where ${x_{i,k}}: \mathcal{R}_{+} \to \mathcal{R}$, $u_i: \mathcal{R}_{+} \to \mathcal{R}$, $y_i: \mathcal{R}_{+} \to \mathcal{R}$ are the states, control input and output of the $i$th agent, respectively, ${f_{i,k}}: \mathcal{R}^{k} \to \mathcal{R}$, $k=1,\cdots,n$ are unknown smooth nonlinear functions.

The mismatched nonlinearities ${f_{i,k}}\left( {{x_{i,1}}, \cdots ,{x_{i,k}}} \right)$ are approximated by radial basis function neural network (RBFNN) over a suitable compact set ${\Omega_\chi} \in{\mathcal{R} ^\iota}$, that is
\begin{flalign}
&f_{i,k}(\chi_{i,k})=W_{i,k}^{T}\phi_{i,k}(\chi_{i,k}) +\varepsilon_{i,k}{(\chi_{i,k})}& \label{eq:16}
\end{flalign}
where $\chi_{i,k}\in{\mathcal{R}^\iota}$ is the NN input vector, $W_{i,k}\in \mathcal{R}^p$ is the ideal weight matrix, $\phi_{i,k}(\chi_{i,k})=\left[\phi_{i,k1}(\chi_{i,k}),\cdots,\phi_{i,kp}(\chi_{i,k})\right]^{T}\in \mathcal{R}^p$ is the basis function vector, and  $\varepsilon_{i,k}{(\chi_{i,k})}\in{\mathcal{R}}$ is the approximate error.
%satisfying  $\left\| {\phi_{i,k}(\chi_{i,k})} \right\|\le {\phi_{m}}$, $\left| {\varepsilon_{i,k} \left(\chi_{i,k} \right)} \right| \le {\varepsilon _m}$, where ${\phi_{m}}$ and ${\varepsilon _m}$ are unknown positive constants.
The typical choice of $\phi_{i,kh}(\chi_{i,k})$, $h=1,\cdots,p$ is the Gaussian function $\phi_{i,kh}(\chi_{i,k})= \exp{\left[-{{(\chi_{i,k}-C_h)^{T} (\chi_{i,k}-C_h)}}/{{b_h^2}}\right]}$, where $C_h= [C_{h1}, \cdots, {C_{h\iota}}]^{T}$ is the center of receptive field, $b_h$ is the width of the Gaussian function.

The objective of this paper is to develop a fully distributed neuroadaptive control scheme for system (\ref{eq:1}) by using intermittent state feedback such that
\begin{itemize}
\item [{$\bullet$}]{All subsystem outputs closely track the desired trajectory $y_0(t)$, with the output tracking error converging to a residual set around zero.}
\item [{$\bullet$}]{All signals in the closed-loop system are semi-globally uniformly ultimately bounded (SGUUB).}
\item [{$\bullet$}]{The Zeno behavior is excluded.}
\end{itemize}

To this end, the following assumptions are needed.

${\textbf{Assumption 1}}$. The directed graph $\mathcal{G}$ is balanced and weakly connected.

${\textbf{Assumption 2}}$.
The full knowledge of $y_0(t)$ is directly accessible by at least one subsystem, i.e., $\sum\nolimits_{i = 1}^N {{\mu_i}>0}$, and $\dot{y}_0(t)$ satisfies $\left| {{{\dot y}_0}(t)} \right| \le {Y_0}$, where $Y_0>0$ is an unknown constant.

${\textbf{Assumption 3}}$.
The basis function vector ${\phi_{i,k}(\chi_{i,k})}$ and the approximate error $\varepsilon_{i,k}{(\chi_{i,k})}$ satisfy $\left\| {\phi_{i,k}(\chi_{i,k})} \right\|\le {\phi_{i,m}}$, $\left| {\varepsilon_{i,k} \left(\chi_{i,k} \right)} \right| \le {\varepsilon _{i,m}}$, respectively, where ${\phi_{i,m}}$ and ${\varepsilon _{i,m}}$ are unknown positive constants.

The following lemma is introduced, which is crucial in the system stability analysis.

${\textbf{Lemma 1}}$ \cite{wang2021adaptive}.
Let $Q =\mathcal{L}+\mathcal{B} +(\mathcal{L} +\mathcal{B})^{T}$. Based on \emph{Assumption} 1, the matrix $Q$ is symmetric and positive definite.

${\textbf{Remark 1}}$.
\emph{Assumption} 1 is quite common in literature \cite{8319522,wang2021adaptive}. Different from the results in \cite{8319522,wang2021adaptive}, as noted in \emph{Assumption} 2, only the available information of ${y_0}(t)$ and its first derivative is needed, but not the high order derivatives. For \emph{Assumption} 3, since the basis function vector $\phi_{i,k}$ of the RBNFF chosen in this paper (i.e., Gaussian function) is inherently bounded, and thus it is valid to assume that $\phi_{i,k}$ is bounded.

\section{Distributed Control Using Continuous State Feedback}
To facilitate the distributed event-triggered control design, we first develop a distributed continuous control scheme under regular state feedback. The tracking errors are defined as follows:
\begin{flalign}
{z_{i,1}}=\,& {x_{i,1}} - {{\hat y}_{i,0}} ={\varepsilon _i} + {{\tilde y}_{i,0}} & \label{eq:8}\\
{z_{i,k}}=\,& {x_{i,k}} - {\alpha _{i,kf}},k = 2, \cdots ,n - 1 & \label{eq:9}\\
{e_i}=\,& \sum\limits_{j = 1}^N {{a_{ij}}} \left( {{y_i} - {y_j}} \right) + {\mu _i}\left( {{y_i} - {y_0}} \right)& \label{eq:10}
\end{flalign}
where ${\varepsilon_i} = {x_{i,1}} - {y_0}$ is the output tracking error. %In each subsystem, ${\hat y_{i,0}}$ is used to estimate of the unknown reference signal $y_0$ with ${\mu _i}=0$, and the information of $y_0$ is accessible by at least one subsystem with $\mu_i=1$.
In each subsystem, $\mu_{i}=1$ is used to indicate the case that the reference signal $y_{0}$ is accessible directly to subsystem $i$; otherwise, $\mu_{i}=0$. For the latter case, we introduce  ${{{\hat y}}_{i,0}}$ to estimate $y_0$, with ${\tilde y_{i,0}}={y_0}-{\hat y_{i,0}}$, as seen later.

As a key design step for control design, we introduce the following first-order filter, which plays a crucial role in the stability analysis:
\begin{flalign}
&{\mu _{i,k}}{\dot \alpha _{i,kf}} +{\alpha _{i,kf}} ={\alpha _{i,k - 1}},\,\,{\alpha _{i,kf}}\left( 0 \right) = {\alpha _{i,k - 1}}\left( 0 \right)& \label{eq:11}
\end{flalign}
for $k=2,\cdots,n$, where ${\mu _{i,k}}>0$ is a time constant, $\alpha_{i,k-1}$ is the virtual control serving as the input of (\ref{eq:11}), and $\alpha_{i, kf}$ is the output. Furthermore, we define
\begin{flalign}
&{y_{i,k}} = {\alpha _{i,kf}} - {\alpha _{i,k - 1}}, k=2,\cdots,n& \label{eq:12}
\end{flalign}
which benefits the stability analysis, as seen later.

The distributed continuous control scheme under regular state feedback is designed as follows:
\begin{flalign}
{\alpha _{i,1}} =\, & - {\kappa _1}{e_i} - \left({c_{i,1}}+1\right) {z_{i,1}} - \hat W_{i,1}^T{\phi _{i,1}}\left( {{\chi _{i,1}}} \right) + {{\dot {\hat y}}_{i,0}}& \label{eq:21}\\
{\alpha _{i,k}} =\, & - (c_{i,k} + 1){z_{i,k}} - {z_{i,k - 1}} {\rm{-}} \hat W_{i,k}^T{\phi _{i,k}}\left( {{\chi _{i,k}}} \right) + {{\dot \alpha }_{i,kf}}&\label{eq:45}\\
{u_i}=\, & {\alpha _{i,n}}& \label{eq:53}
\end{flalign}
for $k=2,\cdots,n$, where $\kappa _1$ and $c_{i,k}$, $k=1,\cdots,n$ are positive design parameters. The parameter estimator ${\dot{{\hat W}}_{i,k}}$ and the distributed estimator ${\dot{\hat y}_{i,0}}$ are designed as:
\begin{flalign}
{{\dot {\hat W}}_{i,k}} =&  - {\Gamma _{i,k}}{\sigma _{i,k}}{{\hat W}_{i,k}} + {\Gamma _{i,k}}{\phi _{i,k}}\left( {{\chi _{i,k}}} \right){z_{i,k}}&\label{eq:22}\\
{{\dot {\hat y}}_{i,0}} =&  - {\gamma _{yi0}}{e_i} - {\gamma _{yi0}}{\sigma _{yi0}}{{\hat y}_{i,0}}&\label{eq:23}
\end{flalign}
for $k=1,\cdots,n$, where ${\sigma _{i,k}}$, ${\gamma _{yi0}}$ and ${\sigma _{yi0}}$ are positive design parameters, $\hat{W}_{i,k}$ is the estimate of ${W}_{i,k}$, with ${\tilde W}_{i,k}={W}_{i,k}- \hat{W}_{i,k}$, and $\Gamma _{i,k}$ is a positive definite design matrix.

Now we are ready to state the following theorem.

${\textbf{Theorem 1}}$.
Consider a strict-feedback nonlinear multiple agent system of $N$ agents (\ref{eq:1}) with a desired trajectory $y_0(t)$ under the directed topology $\mathcal{G}$, satisfying \emph{Assumptions} 1-2, if using distributed neuroadaptive controller (\ref{eq:53}), with the adaptive law (\ref{eq:22}) and the distributed estimator (\ref{eq:23}), then it holds that: i) all signals in the closed-loop system are SGUUB; and ii) the output tracking error ${\varepsilon}= [\varepsilon_1,\cdots, \varepsilon_N]^{T}$ converges to a residual set around zero.

${\textbf{Proof}}$. See Appendix A.

%${\textbf{Remark 3}}$.
%Compared to the traditional centralized tracking problem of single systems \cite{zhang2021adaptive,7355294}, the main challenge in solving the distributed consensus tracking problem, lies in the constraint that only a small fraction of the subsystems can acquire the desired signal information directly. Inspired by the ideas in \cite{wang2021adaptive}, the distributed estimator defined in (\ref{eq:23}) (or (\ref{eq:64})) is introduced to estimate $y_0$ in each subsystem with ${\mu _i}=0$. Therefore, the problem studied here is more general than that in \cite{wang2014distributed,yu2012adaptive}, where $y_0$ is linearly parameterized and the basis function vectors are known by all subsystems.

\section{Distributed Control using intermittent state feedback}
In this section, a fully distributed neuroadaptive backstepping control scheme with directed communication and state-triggering setting is presented.

{\subsection{Event Triggering Mechanism}}
Denote ${{\bar x}_{i,k}}$ and ${{\bar x}_{j,k}}$, $i\in \{{0,\mathcal{V}}\}$, $j \in \mathcal{N}_i$ as the local states information (or the leader, i.e., $i=0$) and its neighboring states information, respectively, which are broadcast to their connected subsystems according to the designed triggering mechanism. Since $t_{k,l}^i$ and $t_{k,l}^j$ denote the $l$th event time for agent $i$ and its neighbor $j$ broadcasting their state information, respectively, it holds that the states of the agent $i$ and its neighbor $j$ are kept unchanged as ${{\bar x}_{i,k}}\left( t \right) ={x_{i,k}}\left( {t_{k,l}^i} \right)$, $\forall t \in [t_{k,l}^i,t_{k,l + 1}^i)$, and ${{\bar x}_{j,k}}\left( t \right) ={x_{j,k}}\left( {t_{k,l}^j} \right), \forall t \in [t_{k,l}^j,t_{k,l + 1}^j)$.

The triggering conditions are then designed as:
\begin{flalign}
& t_{k,l + 1}^i= \inf \left\{ {t > t_{k,l}^i,\left| {{x_{i,k}}\left( t \right) - {{\bar x}_{i,k}}\left( t \right)} \right| > \Delta x_k^i} \right\} & \label{eq:2}\\
& t_{k,l + 1}^j= \inf \left\{ {t > t_{k,l}^j,\left| {{x_{j,k}}\left( t \right) - {{\bar x}_{j,k}}\left( t \right)} \right| > \Delta x_k^j} \right\} & \label{eq:z2}
\end{flalign}
where $\Delta x_k^i$ and $\Delta x_k^j$ are positive  triggering thresholds, $t_{k,0}^i$ is the first instant when (\ref{eq:2}) is fulfilled for subsystem $i$ (or the leader, i.e., $i=0$), and $t_{k,0}^j$ is the first triggering time for its neighbor $j$, $l = 0,1,2, \cdots$, $k=1, \ldots, n$.
\begin{figure}
\begin{center}
\includegraphics[width=0.48\textwidth,height=45mm]{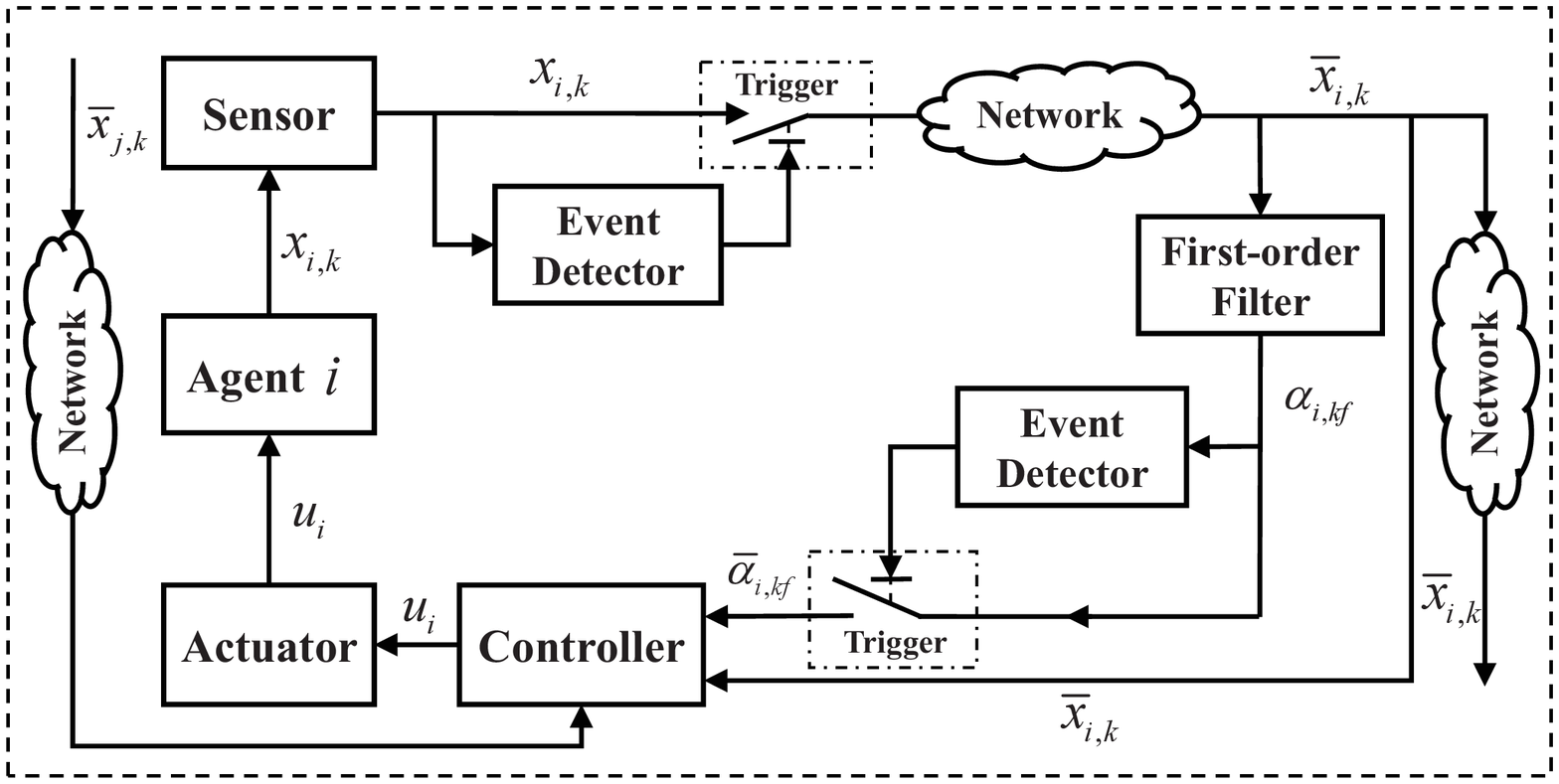}
\caption{A block diagram of the proposed control strategy for agent $i$ with state-triggering setting.}
\end{center}
\end{figure}

{\subsection{Controller Design}}
Firstly, we modify the tracking errors defined in (\ref{eq:8})-(\ref{eq:10}) to the following form:
\begin{flalign}
{\bar{z}_{i,1}}&= {\bar{x}_{i,1}} - {{\hat y}_{i,0}}= {{\varepsilon}_i} + {{\tilde y}_{i,0}} &\label{eq:57}\\
{\bar{z}_{i,k}} &= {\bar{x}_{i,k}} -\bar{\alpha} _{i,kf},k = 2, \cdots ,n - 1 & \label{eq:58}\\
\bar{e}_i &= \sum\limits_{j = 1}^N {{a_{ij}}} \left( {{\bar{y}_i} - {\bar{y}_j}} \right) + {\mu _i}\left( {{\bar{y}_i} - {\bar{y}_0}} \right)& \label{eq:59}
\end{flalign}
where ${{\varepsilon}_i} = {\bar{x}_{i,1}} - {\bar{y}_0}$, ${\tilde y_{i,0}} = {\bar{y}_0} - {\hat y_{i,0}}$, and ${{\bar \alpha }_{i,kf}}\left( t \right)= {\alpha _{i,kf}}\left({{t_{1k,l}^i}} \right)$, $\forall t \in [t_{1k,l}^i,t_{1k,l + 1}^i)$, $k=2,\cdots,n$, with
\begin{flalign}
t_{1k,l + 1}^i {\rm{=}} &\inf \left\{ {t > t_{1k,l}^i,\left| {{\alpha _{i,kf}}\left( t \right) {\rm{-}} {{\bar \alpha }_{i,kf}}\left( t \right)} \right| {\rm{>}}\Delta {\alpha _{kf}^{i}}} \right\}& \label{eq:aa6}
\end{flalign}
where $\Delta {\alpha _{kf}^{i}}$, $k=2,\cdots,n$, is the positive triggering threshold, and $t_{1k,0}^i$ is the first triggering time when (\ref{eq:aa6}) is fulfilled.

Based upon the intermittent state feedback, we redesign the fully distributed event-triggered control scheme as follows:
\begin{flalign}
{{\bar {\alpha}}_{i,1}} =\,& -{\kappa _1}{{\bar e}_i}- \left({c_{i,1}}+1\right){{\bar z}_{i,1}} - \hat W_{i,1}^T{\phi _{i,1}}\left( {{{\bar \chi }_{i,1}}} \right) + {{\dot {\hat y}}_{i,0}}& \label{eq:61}\\
{{\bar \alpha }_{i,k}}=\,& - \left({c_{i,k}}+1\right) {{\bar z}_{i,k}} - {{\bar z}_{i,k - 1}}- \hat W_{i,k}^T{\phi _{i,k}}\left( {{{\bar \chi }_{i,k}}} \right)&  \nonumber \\
&+ \frac{{{{\bar \alpha }_{i,k - 1}} - {{\bar \alpha }_{i,kf}}}}{{{\mu _{i,k}}}},\,k=2,\cdots,n& \label{eq:62}\\
{u_i} =\,& {{\bar \alpha}_{i,n}}& \label{eq:60}
\end{flalign}
where $\kappa _1$ and $c_{i,k}$, $k=1,\cdots,n$ are positive design parameters. The parameter estimator ${\dot{{\hat W}}_{i,k}}$ and the distributed estimator ${\dot{\hat y}_{i,0}}$ are designed as:
\begin{flalign}
{{\dot {\hat W}}_{i,k}} = & - {\Gamma _{i,k}}{\sigma _{i,k}}{{\hat W}_{i,k}} + {\Gamma _{i,k}}{\phi _{i,k}}\left( {{{\bar \chi }_{i,k}}} \right){{\bar z}_{i,k}}& \label{eq:63}\\
{{\dot {\hat y}}_{i,0}} = & - {\gamma _{yi0}}{{\bar e}_i} - {\gamma _{yi0}}{\sigma _{yi0}}{{\hat y}_{i,0}}& \label{eq:64}
\end{flalign}
for $k=1,\cdots,n$, where ${\sigma _{i,k}}$, ${\gamma _{yi0}}$ and ${\sigma _{yi0}}$ are positive design parameters, $\hat{W}_{i,k}$ is the estimate of ${W}_{i,k}$, and $\Gamma _{i,k}$ is a positive definite design matrix.

The proposed distributed control strategy with state-triggering setting is conceptually shown in Fig. 1.

${\textbf{Remark 2}}$.
Here, we must emphasize and clarify that, for the system with state-triggering setting, only the triggered states $\bar{x}_{i,k}\,(k=1,\cdots,n)$ are available for the control design. If the standard backstepping design procedure is adopted, the virtual controllers $\bar{\alpha}_{i,k}(k=2,\cdots,n)$ in (\ref{eq:62}) should be ${{\bar \alpha }_{i,k}}= - \left({c_{i,k}}+1\right) {{\bar z}_{i,k}} -{{\bar z}_{i,k - 1}}- \hat W_{i,k}^T{\phi _{i,k}}( {{{\bar \chi }_{i,k}}})+{\dot{{\bar \alpha }}_{i,{k-1}}}$. Note that the intermittent states $\bar{x}_{i,k}$ are used in ${{{\bar \alpha }}_{i,{k-1}}}$, it holds that ${{\dot{\bar \alpha }}_{i,{k-1}}}$ is discontinuous and unavailable for the backstepping control design.
To overcome this obstacle, in this section we utilize the tracking errors $z_{i,k}$ in (\ref{eq:8})-(\ref{eq:9}) and the virtual control functions $\alpha_{i,k-1}$ in (\ref{eq:21})-(\ref{eq:45}) for the Lyapunov stability analysis.
Nevertheless, it should be noted that $z_{i,k}$ and $\alpha_{i,k-1}$ are not for the controller design, since the final control $v_i$, the parameter updating law ${\dot{\hat{W}}}_{i,k}$ and the distributed estimator ${{\dot {\hat y}}_{i,0}}$ designed in (\ref{eq:60})-(\ref{eq:64}) only utilize intermittent states $\bar{x}_{i,k}$. Such a treatment plays a crucial role in handling the adverse effects arising from state-triggering, as detailed in the sequel.

As the proposed distributed state triggering control is constructed by replacing the states with their triggered ones, it is important to show that such replacement not only saves communication and energy resources but also preserves consensus stability. To this end, we establish the following lemmas.

${\textbf{Lemma 2}}$.
For all $i = 1, \cdots ,N,k = 1, \cdots n$, there exist bounds $\Delta {\phi _{i,k}}>0$, such that
\begin{flalign}
&\left\| {{\phi _{i,k}}\left( {{\chi _{i,k}}} \right) - {\phi _{i,k}}\left( {{{\bar \chi }_{i,k}}} \right)} \right\| \le \Delta {\phi _{i,k}}&\label{eq:b3}
\end{flalign}
where $\Delta {\phi _{i,k}}>0$ is a constant that depends on triggering thresholds $\Delta y_0$, $\Delta x_k^i$, $\Delta x_k^{j}$ and design parameters $b_h$, $j \in \mathcal{N}_i$, $h=1,\cdots,p$.

{\textbf{Proof}}. See Appendix B.

${\textbf{Lemma 3}}$.
The effects of state-triggering are bounded as follows:
\begin{flalign}
\left| {{z_{i,1}} - {{\bar z}_{i,1}}} \right| \le\, & \Delta {x_1^{i}}\buildrel \Delta \over = \Delta{z_{i,1}}& \label{eq:bb3} \\
\left| {{z_{i,r}} - {{\bar z}_{i,r}}} \right| \le \, &\Delta {x_r^{i}} + \Delta {\alpha _{rf}^{i}}\buildrel \Delta \over = \Delta{z_{i,r}},\,r=2,\cdots,n & \label{eq:bb33} \\
\left| {{\alpha _{i,k}} {\rm{-}} {{\bar \alpha }_{i,k}}} \right| \le \,& \sum\limits_{q = 1}^k {{\rho _{i,kq}}\left\| {{{\tilde W}_{i,q}}} \right\|}  + {\tau _{i,k}}\buildrel \Delta \over = \Delta {\alpha _{i,k}} &\label{eq:b1}
\end{flalign}
for $k=1,\cdots,n$, where $\Delta {z_{i,k}}$, ${\rho _{i,kq}}$ and ${\tau _{i,k}}$ are positive constants that depend on the triggering thresholds $\Delta y_0$, $\Delta x_k^i$, $\Delta x_k^j$, $\Delta \alpha _{rf}^i$, topology parameters $d_i$, $\mu_{i}$, design parameters $\kappa_1$, $\mu_{i,r}$, $c_{i,k}$ and $b_{h}$, $r=2,\cdots,n$, $h=1,\cdots,p$.
% $i=1,\cdots,N$,}} $j\in \mathcal{N}_i$, q代替r

{\textbf{Proof}}. See Appendix C.

With those lemmas, we establish the following result.

${\textbf{Theorem 2}}$.
Consider a strict-feedback nonlinear multiple agent system of $N$ agents (\ref{eq:1}) with a desired trajectory $y_0(t)$ under the directed topology $\mathcal{G}$, satisfying \emph{Assumptions} 1-2, if using distributed neuroadaptive controller (\ref{eq:60}), with adaptive law (\ref{eq:63}) and distributed estimator (\ref{eq:64}) under triggering conditions (\ref{eq:2}), (\ref{eq:z2}) and (\ref{eq:aa6}), it then holds that: i) all signals in the closed-loop system are SGUUB; ii) the output tracking error ${\varepsilon}= [\varepsilon_1,\cdots, \varepsilon_N]^{T}$ converges to a residual set around zero, and the upper bound for ${\left\|\varepsilon \right\|_{[0, T]}}$ can be decreased with some proper choices of the design parameters; and iii) the Zeno phenomenon is precluded.
\\
\indent
\textbf{Proof}. The proof involves three parts: stability analysis, performance analysis and exclusion of Zeno behavior.
\\
\indent
\textbf{1) Stability analysis}. This part is comprised of the following $n$ steps.

$\textbf{Step 1}$:
Consider a Lyapunov function ${V_1}= \sum\nolimits_{i = 1}^N \frac{1}{2}{z_{i,1}^2} + \sum\nolimits_{i = 1}^N\frac{1}{2}{\tilde W_{i,1}^T\Gamma _{i,1}^{ - 1}{{\tilde W}_{i,1}}} + \sum\nolimits_{i = 1}^N \frac{1}{2}{y_{i,2}^2} + \sum\nolimits_{i = 1}^N {\frac{{{\kappa _1}}}{{{2\gamma _{yi0}}}}\tilde y_{i,0}^2}$.
From (\ref{eq:1}), (\ref{eq:8}), (\ref{eq:9}), (\ref{eq:12}), (\ref{eq:21}) and using Young's inequality, the derivative of ${V_1}$ is computed as
\begin{flalign}
{{\dot V}_1} \le & - \sum\limits_{i = 1}^N {{\kappa _1}{e_i}{z_{i,1}}}- \sum\limits_{i = 1}^N {{c_{i,1}}z_{i,1}^2} + \sum\limits_{i = 1}^N {\frac{{{\kappa _1}}}{{{\gamma _{yi0}}}}{{\tilde y}_{i,0}}{{\dot {\tilde y}}_{i,0}}} &\nonumber \\
&+ \sum\limits_{i = 1}^N  \left({\tilde W_{i,1}^T\left( {{\phi _{i,1}}\left( {{\chi _{i,1}}} \right){z_{i,1}} - \Gamma _{i,1}^{ - 1}{{\dot {\hat W}}_{i,1}}} \right)}+{l_{i,1}^2}\right) &\nonumber\\
&  - \sum\limits_{i = 1}^N {\mu _{i,2}^*y_{i,2}^2}  + \sum\limits_{i = 1}^N {{z_{i,1}}{z_{i,2}}}  + \sum\limits_{i = 1}^N\frac{1}{2}\varepsilon _{i,m}^2 & \label{eq:66}
\end{flalign}
where ${l_{i,1}} =  - {\dot \alpha _{i,1}}$, $\mu_{i,2}^*$ is a positive design parameter, with $1/{\mu_{i,2}}\ge \mu_{i,2}^*+ 3/4$.

$\textbf{Step \emph{k}}$ $(k=2,\cdots,n-1)$:
Consider a Lyapunov function ${V_k} = {V_{k - 1}} + \sum\nolimits_{i = 1}^N\frac{1}{2} {z_{i,k}^2} + \sum\nolimits_{i = 1}^N\frac{1}{2} {\tilde W_{i,k}^T\Gamma _{i,k}^{ - 1}{{\tilde W}_{i,k}}} + \sum\nolimits_{i = 1}^N\frac{1}{2} {y_{i,k + 1}^2}$.  From (\ref{eq:1}), (\ref{eq:9}), (\ref{eq:12}), (\ref{eq:45}), (\ref{eq:66}) and using Young's inequality, ${\dot{V}_k}$ is expressed as
\begin{flalign}
{{\dot V}_k} \le & - \sum\limits_{i = 1}^N {{\kappa _1}{e_i}{z_{i,1}}}  - \sum\limits_{i = 1}^N \sum\limits_{r = 1}^k {{c_{i,r}}z_{i,r}^2} + \sum\limits_{i = 1}^N {\frac{{{\kappa _1}}}{\gamma _{yi0}}{{\tilde y}_{i,0}}{{\dot {\tilde y}}_{i,0}}}  & \nonumber\\
&+ \sum\limits_{i = 1}^N \sum\limits_{r = 1}^k \left({\tilde W_{i,r}^T\left( {{\phi _{i,r}}\left( {{\chi _{i,r}}} \right){z_{i,r}} {\rm{-}} \Gamma _{i,r}^{ - 1}{{\dot {\hat W}}_{i,r}}} \right)} + {l_{i,r}^2}\right) & \nonumber\\
&{\rm{-}} \sum\limits_{i = 1}^N {\sum\limits_{r = 1}^k {\mu _{i,r + 1}^*y_{i,r + 1}^2} }{\rm{+}} \sum\limits_{i = 1}^N {{z_{i,k}}{z_{i,k + 1}}} + \sum\limits_{i = 1}^N\frac{k}{2}\varepsilon _{i,m}^2
& \label{eq:68}
\end{flalign}
where ${l_{i,k}}=- {\dot \alpha _{i,k}}$, $\mu_{i,k+1}^*$ is a positive design parameter, with $1/{\mu_{i,k+1}}\ge \mu_{i,k+1}^*+ 3/4$.

$\textbf{Step \emph{n}}$:
Consider a Lyapunov function ${V_n} = {V_{n - 1}} + \sum\nolimits_{i = 1}^N\frac{1}{2} {z_{i,n}^2} + \sum\nolimits_{i = 1}^N\frac{1}{2} {\tilde W_{i,n}^T\Gamma _{i,n}^{ - 1}{{\tilde W}_{i,n}}}$.  From (\ref{eq:1}), (\ref{eq:9}) and (\ref{eq:68}), it holds that
\begin{flalign}
{{\dot V}_n} \le &- \sum\limits_{i = 1}^N {\kappa _1}{e_i}{z_{i,1}} - \sum\limits_{i = 1}^N {\sum\limits_{r = 1}^{n - 1} {{c_{i,r}}z_{i,r}^2} } + \sum\limits_{i = 1}^N {\frac{{{\kappa _1}}}{{{\gamma _{yi0}}}}{{\tilde y}_{i,0}}{{\dot {\tilde y}}_{i,0}}} & \nonumber \\
&+ \sum\limits_{i = 1}^N {{z_{i,n}}\left( {{u_i} + W_{i,n}^T{\phi _{i,n}}\left( {{\chi _{i,n}}} \right) - {{\dot \alpha }_{i,nf}} + {z_{i,n}}} \right)}  & \nonumber\\
&+ \sum\limits_{i = 1}^N \sum\limits_{r = 1}^{n - 1} {\tilde W_{i,r}^T\left( {{\phi _{i,r}}\left( {{\chi _{i,r}}} \right){z_{i,r}} - \Gamma _{i,r}^{ - 1}{{\dot {\hat W}}_{i,r}}} \right)}  & \nonumber\\
&-\sum\limits_{i = 1}^N {\tilde W_{i,n}^T\Gamma _{i,n}^{ - 1}{{\dot {\hat W}}_{i,n}}}- \sum\limits_{i = 1}^N {\sum\limits_{r = 1}^{n - 1} {\mu _{i,r + 1}^*y_{i,r + 1}^2} }& \nonumber\\
& + \sum\limits_{i = 1}^N {\sum\limits_{r = 1}^{n - 1} {l_{i,r}^2} }+ \sum\limits_{i = 1}^N z_{i,{n-1}}z_{i,n} + \sum\limits_{i = 1}^N\frac{{2n {\rm{-}} 1}}{4}\varepsilon _{i,m}^2. & \label{eq:73}
\end{flalign}
From (\ref{eq:64}), it is seen that $\sum\nolimits_{i = 1}^N \frac{{{\kappa _1}}}{{{\gamma _{yi0}}}}{{\tilde y}_{i,0}}{{\dot {\tilde y}}_{i,0}} -\sum\nolimits_{i = 1}^N {\kappa _1}{e_i}{z_{i,1}} \le \sum\nolimits_{i = 1}^N {\kappa _1}{\sigma _{yi0}}{{\tilde y}_{i,0}}{{\hat y}_{i,0}}+\sum\nolimits_{i = 1}^N \frac{{{\kappa _1}}}{{{\gamma _{yi0}}}}\left| {{{\tilde y}_{i,0}}} \right|\left| {{{\dot {\bar y}}_0}} \right|+\sum\nolimits_{i = 1}^N {\kappa _1}\left|{{\tilde y}_{i,0}}\right|\Delta {e_i} - \sum\nolimits_{i = 1}^N {\kappa _1}{e_i}{\varepsilon _i}$, where $\Delta {e_i}$ is defined in (\ref{eq:a1})). By using Young's inequality, one has
\begin{flalign}
\sum\limits_{i = 1}^N{\kappa _1}{\sigma _{yi0}}{{\tilde y}_{i,0}}{{\hat y}_{i,0}} \le &
- \sum\limits_{i = 1}^N\frac{{3{\kappa _1}{\sigma _{yi0}}}}{4}\tilde y_{i,0}^2 + \sum\limits_{i = 1}^N{\kappa _1}{\sigma _{yi0}}y_0^2 & \label{eq:0706_1}\\
\sum\limits_{i = 1}^N\frac{{{\kappa _1}}}{{{\gamma _{yi0}}}}\left| {{{\tilde y}_{i,0}}} \right|\left| {{{\dot {\bar y}}_0}} \right|
\le &\sum\limits_{i = 1}^N\frac{{{\kappa _1}{\sigma _{yi0}}}}{4}\tilde y_{i,0}^2 + \sum\limits_{i = 1}^N\frac{{{\kappa _1}}}{{\gamma _{yi0}^2{\sigma _{yi0}}}}{{Y}_0^2} & \label{eq:0706_2}\\
\sum\limits_{i = 1}^N{\kappa _1}\left|{{\tilde y}_{i,0}}\right|\Delta {e_i} \le &
\sum\limits_{i = 1}^N\frac{{{\kappa _1}{\sigma _{yi0}}}}{4}\tilde y_{i,0}^2 + \sum\limits_{i = 1}^N\frac{{{\kappa _1}}}{{{\sigma _{yi0}}}}\Delta e_i^2. & \label{eq:75}
\end{flalign}
In view of (\ref{eq:73}), (\ref{eq:0706_1}), (\ref{eq:0706_2}) and (\ref{eq:75}), it holds that
%\begin{flalign}
%{{\dot V}_n} \le & - \sum\limits_{i = 1}^N {\frac{{{\kappa _1}{\lambda _{\min }}\left( Q \right)}}{2}{{\left\| \varepsilon  \right\|}^2}}  - \sum\limits_{i = 1}^N {\sum\limits_{r = 1}^{n - 1} {{c_{i,r}}z_{i,r}^2} }  & \nonumber \\
%& + \sum\limits_{i = 1}^N \sum\limits_{r = 1}^{n - 1} {\tilde W_{i,r}^T\left( {{\phi _{i,r}}\left( {{\chi _{i,r}}} \right){z_{i,r}} - \Gamma _{i,r}^{ - 1}{{\dot {\hat W}}_{i,r}}} \right)} & \nonumber\\
%&+ \sum\limits_{i = 1}^N {{z_{i,n}}\left( {{u_i} + W_{i,n}^T{\phi _{i,n}}\left( {{\chi _{i,n}}} \right) {\rm{-}} {{\dot \alpha }_{i,nf}} + {z_{i,n}}} \right)} & \nonumber\\
%&- \sum\limits_{i = 1}^N {\sum\limits_{r = 1}^{n - 1} {\mu _{i,r + 1}^*y_{i,r + 1}^2} }  {\rm{-}} \sum\limits_{i = 1}^N {\tilde W_{i,n}^T\Gamma _{i,n}^{ - 1}{{\dot {\hat W}}_{i,n}}}  {\rm{+}} M & \nonumber\\
%& - \sum\limits_{i = 1}^N {\frac{{{\kappa _1}{\sigma _{yi0}}}}{4}\tilde y_{i,0}^2}+ \sum\limits_{i = 1}^N {\sum\limits_{r = 1}^{n - 1} {l_{i,r}^2} } {\rm{+}} \sum\limits_{i = 1}^N z_{i,{n-1}}z_{i,n}
%& \label{eq:76}
%\end{flalign}
\begin{flalign}
{{\dot V}_n} \le & - {\frac{{{\kappa _1}{\lambda _{\min }}\left( Q \right)}}{2}{{\left\| \varepsilon  \right\|}^2}}- \sum\limits_{i = 1}^N {\sum\limits_{r = 1}^{n - 1} {{c_{i,r}}z_{i,r}^2} }{\rm{-}} \sum\limits_{i = 1}^N {\frac{{{\kappa _1}{\sigma _{yi0}}}}{4}\tilde y_{i,0}^2} & \nonumber \\
& + \sum\limits_{i = 1}^N \sum\limits_{r = 1}^{n - 1} {\tilde W_{i,r}^T\left( {{\phi _{i,r}}\left( {{\chi _{i,r}}} \right){z_{i,r}} {\rm{-}} \Gamma _{i,r}^{ - 1}{{\dot {\hat W}}_{i,r}}} \right)}+ \sum\limits_{i = 1}^N {\sum\limits_{r = 1}^{n - 1} {l_{i,r}^2}}  & \nonumber\\
&+ \sum\limits_{i = 1}^N {{z_{i,n}}\left( {{u_i} + W_{i,n}^T{\phi _{i,n}}\left( {{\chi _{i,n}}} \right) {\rm{-}} {{\dot \alpha }_{i,nf}} + {z_{i,n}}}{\rm{+}}z_{i,{n-1}} \right)} & \nonumber\\
&- \sum\limits_{i = 1}^N {\tilde W_{i,n}^T\Gamma _{i,n}^{ - 1}{{\dot {\hat W}}_{i,n}}} - \sum\limits_{i = 1}^N {\sum\limits_{r = 1}^{n - 1} {\mu _{i,r + 1}^*y_{i,r + 1}^2} } + M
& \label{eq:76}
\end{flalign}
where $M=\sum\nolimits_{i = 1}^N {{{\kappa _1}{\sigma _{yi0}}y_0^2}}  + \sum\nolimits_{i = 1}^N {{\frac{{{\kappa _1}}}{{\gamma _{yi0}^2{\sigma _{yi0}}}}{Y_0}^2}} + \sum\nolimits_{i = 1}^N {\frac{{{\kappa _1}}}{{{\sigma _{yi0}}}}\Delta e_i^2} + \sum\nolimits_{i = 1}^N\frac{{2n - 1}}{4}\varepsilon _{i,m}^2$.
Notice that the actual control law $u_i$ in (\ref{eq:60}) can be rewritten as
\begin{flalign}
{u_i} = & - \left( {{c_{i,n}} + 1} \right){z_{i,n}} - {z_{i,n - 1}} - \hat W_{i,n}^T{\phi _{i,n}}\left( {{\chi _{i,n}}} \right)& \nonumber\\
&+ \frac{{{\alpha _{i,n - 1}} - {\alpha _{i,nf}}}}{{{\mu _{i,n}}}} + \left( {{c_{i,n}} + 1} \right)\left( {{z_{i,n}} - {{\bar z}_{i,n}}} \right)& \nonumber\\
&+ \hat W_{i,n}^T\left( {{\phi _{i,n}}\left( {{\chi _{i,n}}} \right) - {\phi _{i,n}}\left( {{{\bar \chi }_{i,n}}} \right)} \right) + {z_{i,n - 1}}& \nonumber\\
&- {{\bar z}_{i,n - 1}} + \frac{{{{\bar \alpha }_{i,n - 1}} - {\alpha _{i,n - 1}}}}{{{\mu _{i,n}}}} + \frac{{{\alpha _{i,nf}} - {{\bar \alpha }_{i,nf}}}}{{{\mu _{i,n}}}}. &\label{eq:z78}
\end{flalign}
Using (\ref{eq:63}) and (\ref{eq:z78}), then ${{\dot V}_n}$ becomes
\begin{flalign}
{{\dot V}_n} \le&  - {\frac{{{\kappa _1}{\lambda _{\min }}\left( Q \right)}}{2}{{\left\| \varepsilon  \right\|}^2}}-\sum\limits_{i = 1}^N {\sum\limits_{r = 1}^n {{c_{i,r}}z_{i,r}^2} }{\rm{-}} \sum\limits_{i = 1}^N {\frac{{{\kappa _1}{\sigma _{yi0}}}}{4}\tilde y_{i,0}^2} & \nonumber\\
&- \sum\limits_{i = 1}^N {\sum\limits_{r = 1}^n {\frac{{{\sigma _{i,r}}}}{2}\tilde W_{i,r}^T{{\tilde W}_{i,r}}} } + \sum\limits_{i = 1}^N {\Delta {W_{i,\phi }} + } \sum\limits_{i = 1}^N \left|z_{i,n}\right| \Delta {\alpha_{i,n}}  & \nonumber\\
&- \sum\limits_{i = 1}^N {\sum\limits_{r = 1}^{n - 1} {\mu _{i,r + 1}^*}y_{i,r + 1}^2}+\sum\limits_{i = 1}^N {\sum\limits_{r = 1}^{n - 1} {l_{i,r}^2} } + M_1& \label{eq:81}
\end{flalign}
where $\Delta {\alpha_{i,n}}=({c_{i,n}+ 1})\Delta {z_{i,n}}+ \Delta {\alpha _{i,n - 1}/{{\mu _{i,n}}}+ {\Delta {\alpha _{i,nf}}}}/{{\mu _{i,n}}}+ \Delta {z_{i,n - 1}}+\left| {\hat W_{i,n}^T\left( {{\phi _{i,n}}\left( {{\chi _{i,n}}} \right) - {\phi _{i,n}}\left( {{{\bar \chi }_{i,n}}} \right)} \right)} \right|$, ${\Delta {W_{i,\phi }}}=\sum\nolimits_{r = 1}^n\tilde W_{i,r}^T \left( {{\phi _{i,r}}\left( {{\chi _{i,r}}} \right){z_{i,r}} - {\phi _{i,r}}\left( {{{\bar \chi }_{i,r}}} \right){{\bar z}_{i,r}}} \right)$, $\Delta {\alpha_{i,{nf}}}={\left| {{\alpha_{i,{nf}}} - {\bar{\alpha}_{i,{nf}}}} \right|}$ and $M _1=M+\sum\nolimits_{i = 1}^N {\sum\nolimits_{r = 1}^{n}}{\frac{{{\sigma _{i,r}}}}{2}{{\left\| {{W_{i,r}}} \right\|}^2}}$. Invoking \emph{Lemmas} 2-3 and using Young's inequality, it holds that
\begin{flalign}
\left|z_{i,n}\right| \Delta {\alpha_{i,n}} \le & \frac{{{c_{i,n}}}}{2}z_{i,n}^2 + \sum\limits_{r = 1}^n {\frac{{\rho _{i,nr}^2}}{{{c_{i,n}}}}\tilde W_{i,r}^T{{\tilde W}_{i,r}}}  + \frac{\tau _{i,n}^2 }{{{c_{i,n}}}}& \label{eq:aa82}\\
\Delta {W_{i,\phi }} \le &\sum\limits_{r = 1}^n {\left( {\Delta {\phi _{i,r}}\left| {{z_{i,r}}} \right| +\phi _{i,m}}\Delta {z_{i,r}} \right)} \left\| {{{\tilde W}_{i,r}}} \right\|&\nonumber\\
\le &\sum\limits_{r = 1}^n {\frac{{{c_{i,r}}}}{4}z_{i,r}^2}  + \sum\limits_{r = 1}^n \frac{{\Delta \phi _{i,r}^2}} {{{c_{i,r}}}} \tilde W_{i,r}^T{{\tilde W}_{i,r}} & \nonumber\\
&+\frac{1}{4}\tilde W_{i,r}^T{{\tilde W}_{i,r}}+ \sum\limits_{r = 1}^n {{\phi _{i,m}^2}\Delta z_{i,r}^2}.
& \label{eq:a82}
\end{flalign}
From (\ref{eq:aa82}) and (\ref{eq:a82}), it can be derived from (\ref{eq:81}) that
\begin{flalign}
{{\dot V}_n} \le & - {\frac{{{\kappa _1}{\lambda _{\min }}\left( Q \right)}}{2}{{\left\| \varepsilon  \right\|}^2}}  - \sum\limits_{i = 1}^N {\sum\limits_{r = 1}^n {c_{i,r}^*z_{i,r}^2} } & \nonumber\\
&- \sum\limits_{i = 1}^N {\frac{{{\kappa _1}{\sigma _{yi0}}}}{4}\tilde y_{i,0}^2} - \sum\limits_{i = 1}^N {\sum\limits_{r = 1}^n {\frac{{{\sigma _{i,r}^*}}}{2}\tilde W_{i,r}^T{{\tilde W}_{i,r}}} }& \nonumber\\
&-\sum\limits_{i = 1}^N {\sum\limits_{r = 1}^{n - 1} {\mu _{i,r + 1}^*y_{i,r + 1}^2} } +\sum\limits_{i = 1}^N {\sum\limits_{r = 1}^{n - 1} {l_{i,r}^2} }  + {\Gamma_n} & \label{eq:a83}
\end{flalign}
where ${\Gamma_n} =M_1 + \sum\nolimits_{i = 1}^N{\frac{{\tau_{i,n}^2}}{{{c_{i,n}}}}}+\sum\nolimits_{i = 1}^N {\sum\nolimits_{r = 1}^{n}}{\phi _{i,m}^2}\Delta z_{i,r}^2$. Since for any $\rho>0$, the set defined by
${\Omega _v}:=\{\sum\nolimits_{i = 1}^N {\sum\nolimits_{r = 1}^{n}}{z_{i,r}^2}
+\sum\nolimits_{i = 1}^N {\sum\nolimits_{r = 1}^{n}} {\tilde W_{i,r}^T\Gamma _{i,r}^{ - 1}{{\tilde W}_{i,r}}}{\rm{+}} \sum\nolimits_{i = 1}^N {\sum\nolimits_{r = 1}^{n-1}}{y_{i,r + 1}^2}  {\rm{+}} \sum\nolimits_{i = 1}^N\frac{{{\kappa _1}}}{{\gamma _{yi0}}}\tilde y_{i,0}^2$ $\le 2\rho\}$ is a compact one. Thus $\left| {{l_{i,r}}} \right| \le \varpi_{i,r}$ on $\Omega_v$, $r=1,\cdots,n-1$, it follows that
\begin{flalign}
&{{\dot V}_n} \le - {\frac{{{\kappa _1}{\lambda _{\min }}\left( Q \right)}}{2}{{\left\| \varepsilon  \right\|}^2}}-\lambda {V_n} + {\Delta _n}& \label{eq:83}
\end{flalign} %\left\{   \right\}
where $\lambda  = \min \left\{\right. 2{c_{i,1}^*},\cdots,2{c_{i,n}^*}, \frac{{\sigma _{i,1}^*}}{{\lambda _{\max }}\{{\Gamma_{i,1}^{-1}} \}},\cdots, \frac{{\sigma _{i,n}^*}}{{\lambda _{\max }}\{{\Gamma_{i,n}^{-1}}\}}$, $2\mu _{i,2}^*,\cdots,2\mu _{i,n}^*, \frac{1}{2}{{\sigma _{yi0}}{\gamma _{yi0}}} \left. \right\} $, $c_{i,r}^* = \frac{3}{4}{c_{i,r}}$, $r=1,\cdots,n-1$, $c_{i,n}^* = \frac{1}{4}{c_{i,n}}$, $\sigma _{i,k}^* \ge \frac{{{\sigma _{i,k}}}}{2} - \frac{{\Delta \phi_{i,k}^2}} {{{c_{i,k}}}} - \frac{{\rho _{i,nk}^2}} {{{c_{i,n}}}} - \frac{1}{4}$, $k=1,\cdots,n$, and ${\Delta _n} ={\Gamma_n} + \sum\nolimits_{i = 1}^N {\sum\nolimits_{r = 1}^{n-1}}{\varpi_{i,r}^2}$.

In view of (\ref{eq:83}), we have ${{\dot V}_n}\le -\lambda {V_n} + {\Delta _n}$, which implies that ${{\dot V}_n}<0$ on $V_n=\rho$ when $\lambda>\frac{\Delta_n}{\rho}$. Thus $V_n(t)\le \rho$ is an invariant set, i.e., ${V_{n}} \in {L_\infty}$, it follows that the signals ${z_{i,k}} $, ${\tilde W_{i,k}}$, ${y_{i,r + 1}}$, ${\tilde y_{i,0}}$ are bounded, $k = 1, \cdots ,n,\,r = 1, \cdots ,n - 1$. From (\ref{eq:8})-(\ref{eq:10}) and  (\ref{eq:12})-(\ref{eq:45}), it follows that $x_{i,k}$, $k=1,\cdots,n$ is bounded. Then according to \emph{Lemma} 3, it can be derived from (\ref{eq:60}) that $u_i$ is bounded. Therefore, all signals in the closed-loop system are ensured to be SGUUB.

\textbf{2) Performance analysis}.
By the definition of $V_n$ and using (\ref{eq:83}), we have ${\left\| \varepsilon  \right\|^2} \le\sum\nolimits_{i = 1}^N\frac{1}{2}{z_{i,1}^2 + \sum\nolimits_{i = 1}^N\frac{1}{2}\tilde y_{i,0}^2} \le \hbar {V_n}$, it follows that ${\left\| \varepsilon  \right\|^2} \le \hbar \left[ {{e^{ - \lambda t}}{V_n}\left( 0 \right) + \frac{{{\Delta _n}}}{\lambda }\left( {1 - {e^{ - \lambda t}}} \right)} \right]$, where $\hbar = \max \left\{ {1,\frac{{{\gamma _{yi0}}}}{{{\kappa _1}}}} \right\}>0$ is a constant. Thus it can be concluded that the output tracking error ${\varepsilon}=[\varepsilon_1,\cdots,\varepsilon_N]^{T}$  will attenuate to a residual set around zero. In addition, from (\ref{eq:83}), we can obtain that
\begin{flalign}
&\|\varepsilon(t)\|_{[0, T]}\leq \sqrt{\frac{2}{\kappa_1 \lambda_{\min }(Q)}\left[\frac{V_{n}(0)}{T}+\Delta_{n}\right]}
&\label{eq:a57}
\end{flalign}
which implies that the upper bound of  ${\left\|\varepsilon \right\|_{[0, T]}}$ can be decreased by decreasing the triggering thresholds $\Delta y_0$, $\Delta x_k^{i}$, $\Delta x_k^{j}$ and $\Delta \alpha_{rf}^{i}$, and increasing design parameters $c_{i,k}$, $\Gamma _{i,k}$, $\gamma _{yi0}$ and $\mu_{i,r}$, $k=1,\cdots,n$, $r=2,\cdots,n$.

\textbf{3) Exclusion of Zeno behavior}.
Finally, we show that the result iii) is ensured. Define $\omega_{k,l}^{i}(t)=x_{i,k}(t) -\bar{x}_{i,k}(t)$, $\forall t \in\left[t_{k,l}^{i}, t_{k,l+1}^{i} \right)$, $i \in \left\{ {0,\mathcal{V}} \right\}$, $j \in \mathcal{N}_i$, $k=1,\cdots,n$. The derivative of $\omega_{k,l}^{i}$ is computed as
\begin{flalign}
&\frac{d\left|\omega_{k,l}^{i}\right|}{d t}= \frac{d\left(\omega_{k,l}^{i} {\rm{\times}} \omega_{k,l}^{i}\right)^{\frac{1}{2}}}{d t}= \rm{sign}\left(\omega_{k,l}^{i}\right) \dot{\omega}_{k,l}^{i}\leq\left|\dot{\omega}_{k,l}^{i}\right|.
&\label{eq:aaa57}
\end{flalign}
Since $\bar{x}_{i,k}(t)$ remains unchanged for $t \in\left[t_{k,l}^{i}, t_{k,l+1}^{i}\right)$, one obtains that $\left|\dot{\omega}_{k,l}^{i} \right|=\left|{x_{i,k + 1}} + {f_{i,k}}\right|$, $k=1,\cdots,n-1$, and $\left|\dot{\omega}_{n,l}^{i}\right| =\left|{u_i} + {f_{i,n}}\right|$, with $\left|\dot{\omega}_{k,l}^{0}\right| =\left|\dot{y}_{0}\right|$. By the boundedness of $x_{i,k}$, $u_i$ and $f_{i,k}$, it follows that $\left|\dot{\omega}_{k,l}^{i}\right| \leq \omega_{0}^{i}$, $k=1,\cdots,n$, where $\omega_{0}^{i}$ is an unknown positive constant, then we have $t_{k,l+1}^{i}-t_{k,l}^{i} \geq \Delta x_k^i/ \omega_{0}^{i} > T_i$. Similarly, it holds that $t_{r,l+1}^{j}-t_{r,l}^{j} > T_j$ and $t_{1r,l+1}^{i}-t_{1r,l}^{i} > T_{\alpha}$, $r=2,\cdots,n$, where $T_i$, $T_j$ and $T_{\alpha}$ are positive constants. Therefore the Zeno behavior is excluded. The proof is completed.
$\hfill\blacksquare$

%To see the novelty of the proposed control method, following several comments are worth making.
${\textbf{Remark 3}}$.
To see the novelty of the proposed control method, it is worth making the following comments: i) The proposed distributed scheme is based upon the networked strict-feedback systems with directed communication and state-triggering setting, which includes the existing results in \cite{wang2020adaptive,wang2021adaptive,long2022output} as special cases; ii) Different from the methods in \cite{zhang2021adaptive,wang2020adaptive,wang2021adaptive},  where the system uncertainties must be parametric, the agent model considered here is non-identical and in strict-feedback form with mismatched and nonparametric uncertainties; and iii) Unlike the results in \cite{wang2020adaptive,wang2021adaptive}, the communication regarding all states is executed via an event-based discontinuous way in this work, which saves both communication and energy resources.

${\textbf{Remark 4}}$.
Here we pause to stress the following merits associated with the proposed distributed triggering mechanism, from two aspects. Firstly, it is seen from (\ref{eq:2}), (\ref{eq:z2}) and (\ref{eq:aa6})-(\ref{eq:60}) that, no prior knowledge of global communication graph is required in both event-triggered rules and distributed control laws. Hence, the proposed method can be implemented in a fully distributed manner, completely avoiding the requirements on the continuous monitoring of neighbors' states in \cite{6068223,8319522,8910377}. Secondly, in contrast to \cite{wang2020adaptive,wang2021adaptive}, the proposed triggering mechanism allows the communication regarding all states to be executed in an intermitted pace.

${\textbf{Remark 5}}$.
It is worth mentioning how the issue of non-differentiability of virtual control associated with backstepping design arising from intermittent (triggered) state feedback is addressed. For normal form systems, one can use the partial derivatives $\frac{\partial \alpha_{i,k-1}}{\partial x_{i,k}}\,(i=1,\ldots, N, k=2, \ldots, n)$ in deriving the final controller to circumvent the non-differentiability problem of $\bar{\alpha}_{i,k-1}$ in the control design, as seen in \cite{wang2021adaptive}. However, such treatment relies on the key precondition that the partial derivatives $\frac{\partial \alpha_{i,k-1}}{\partial x_{i,k}}$ are constants, thus is only suitable for systems in normal form. Here by using the idea of dynamic filtering technique, we develop a fully distributed neuroadaptive event-triggered control scheme that not only avoids the non-differentiability issue in the backstepping design but also is applicable to nonlinear systems in either normal or strict-feedback form.

${\textbf{Remark 6}}$.
To recap the control design philosophy utilized in this work, it is worth noting that although the calculation of the repetitive differentiation of virtual control signal is averted in the control design by using the dynamic filtering technique, such calculation is still involved in stability analysis, thus the non-differentiability issue associated with virtual controllers cannot be addressed by using the first-order filter directly.
To remove this obstacle, here the backstepping technique is not adopted directly in designing the distributed event-triggered controller $u_i$ and virtual controllers $\bar{\alpha}_{i,k-1}$. Instead, a distributed continuous control scheme is firstly constructed upon regular state feedback, hence the designed virtual controllers ${\alpha}_{i,k-1}$ remain differentiable. Based on such control structure, we further derive a new distributed event-triggered control scheme by replacing the states in the preceding scheme with the triggered ones.

${\textbf{Remark 7}}$.
In deriving the fully distributed event-triggered control scheme (\ref{eq:61})-(\ref{eq:60}), the main challenge in the stability analysis lies in how to deal with the effect caused by triggering errors $x_{i,k}-\bar{x}_{i,k}$, $k=1,\cdots,n$, when the triggered states $\bar{x}_{i,k}$ are utilized in the controller design. In fact, the major difficulty is the treatment of the terms $\sum\nolimits_{i = 1}^N \Delta {W_{i,\phi }}$ and $\sum\nolimits_{i = 1}^N\left|z_{i,n}\right| \Delta {\alpha_{i,n}}$ in (\ref{eq:81}), which are shown to be bounded by the critical results established in \emph{Lemmas} 2-3. Furthermore, it is shown that the bounds rely only on the tracking error $z_{i,k}$ and parameter estimation error ${{\tilde W}_{i,k}}$, and thus can be incorporated into two negative terms $- \sum\nolimits_{i = 1}^N \sum\nolimits_{r = 1}^n {{c_{i,r}^*}z_{i,r}^2} $ and $-\sum\nolimits_{i = 1}^N \sum\nolimits_{r = 1}^n {\frac{{{\sigma _{i,r}^*}}}{2}\tilde W_{i,r}^T{{\tilde W}_{i,r}}}$ of $\dot{V}_n$, as seen in (\ref{eq:a83}). It is such treatment that allows the impact stemming from state-triggering to be gradually handled, ensuring the stability of the closed-loop system.

${\textbf{Remark 8}}$:
It is observed from (\ref{eq:a57}) that the larger the triggering thresholds used, the smaller the triggering times would be, and thus more communication resources are saved. Whereas, the upper bound of $\|\varepsilon(t)\|_{[0, T]}$ is also increased. Also, it is not difficult to see that larger design parameters $c_{i,k}$, $\Gamma _{i,k}$, $\gamma _{yi0}$ and $\mu_{i,r}\,(k=1,\cdots,n,r=2,\cdots,n)$ leads to smaller $\|\varepsilon(t)\|_{[0, T]}$, which allows the proposed scheme to enhance the tracking performance to a certain extent. Nevertheless, it should be pointed out that larger design parameters inevitably leads  to larger control effort. Therefore, a trade-off should be made considering communication cost, tracking performance and control effort in practice.

${\textbf{Remark 9}}$.
To mitigate the communication burden in networked control systems, various types of input-triggered adaptive control laws are developed \cite{xing2016event,8474310,cao2019event}, and those solutions are quite well established. Besides, due to the input-triggered control does not involve any intermittent states, it has no effect on the first $n-1$ steps of the backstepping design procedure. Therefore it is not difficult to design input-triggered control schemes by using backstepping techniques. In this work we propose the event triggering mechanism that triggers both the states from the agent itself (including the leader, i.e., $i=0$) and its neighbors. In this way, the communication among all the states is executed intermittently on an event-driven basis, which allows the communication resources utilized among the agents and the connected subsystems to be substantially saved. To make the description of the technical contents more comprehensible and accessible, input-triggering is not considered in the development. Including such element, however, can be addressed similarly and it is quite straightforward to prove the stability accordingly.

${\textbf{Remark 10}}$.
Inspired by the ideas in \cite{wang2020adaptive,wang2021adaptive,KUMARI2020109163}, the triggering mechanism with fixed constant triggering thresholds is adapted in this work. It contributes preventing the occurrence of Zeno behavior and at the time avoiding the continuous monitoring of neighbors' states in \cite{6068223,8319522,8910377}.  Nevertheless, it should be pointed out that such triggering mechanism reduces the accuracy of transmitted signal to some extent and may further impacts the tracking performance. To solve these problems, numerous adaptive event-triggered algorithms with time-varying triggering thresholds have been developed
\cite{xing2016event,7920382,long2022output}. As our focus in this work is mainly on developing a fully distributed event-triggered control scheme for networked nonlinear strict-feedback systems with mismatched uncertainties, instead of optimizing the triggering mechanism, the problem with time-varying triggering thresholds is of great importance and deserves attention in the future study.

\section{Simulation Verification}
A group of 4 nonlinear subsystems modeled with the following dynamics is considered:
\begin{flalign}
{{\dot x}_{i,1}} =\,& {x_{i,2}} + 0.5\sin \left( {0.1{x_{i,1}}} \right) & \nonumber\\
{{\dot x}_{i,2}} =\, &{u_i} + 0.1\sin \left( {{x_{i,1}}{x_{i,2}}} \right) + 0.2{e^{ - {{\left| {{x_{i,1}}} \right|}^i} + 1}} & \nonumber\\
{y_i} =\,& {x_{i,1}}& \label{eq:zz3}
\end{flalign}
for $i=1,\cdots,4$. Fig. 2 shows the interaction topology of the multiple agent system. In the simulation, we set the desired trajectory $y_0=0.5\sin(0.1t) + 0.5\sin(0.05t)$, the initial states $x_{i,1}(0)=1.0$, $x_{i,2}(0)=0$, the triggering thresholds $\Delta x_1^i=0.01$, $\Delta x_2^i=0.01$, $\Delta \alpha_{2f}^i=0.02$, $\Delta y_0=0.005$, the design parameters $k_1=0.5$, $c_1=5.0$, $c_2=5.0$, $\gamma_{yi0}=1.5$, $\sigma_{yi0}=0.001$, $\sigma_{i,1}=20$,  $\Gamma_{i,1}=0.005$, $\sigma_{i,2}=20$,  $\Gamma_{i,2}=0.005$ and ${\mu _{i,2}}=0.2$. The RBFNN contains 25 nodes with centers distributed in the space [-5,5], and $b_h=2$. The results are presented in Fig. 3. Fig. 3 (a) shows the output trajectories of all subsystems. From Fig. 3 (b), it can be determined that the output tracking error ${{\varepsilon}_i}$ converges to a compact set around origin. Fig. 3 (c) gives the distributed protocol $u_i$. The triggered times of states $x_{i,1}$ and $x_{i,2}$ are presented in Fig. 3 (d).

Furthermore, to test the effect of triggering thresholds on the system tracking performance, we choose $\Delta x_1^i=0.1$ and $\Delta x_2^i=0.1$, and the same set of other design parameters are used. The results are presented in Fig. 3 (e)-(f). The number of triggering events for $x_{i,1}$ and $x_{i,2}$ under different triggering thresholds is shown in Table 1.  The results show that the larger the triggering thresholds are used, the fewer the triggering times are needed. Nevertheless, the output tracking error becomes slightly larger.

\begin{figure}
\begin{center}
\includegraphics[width=0.37\textwidth,height=33mm]{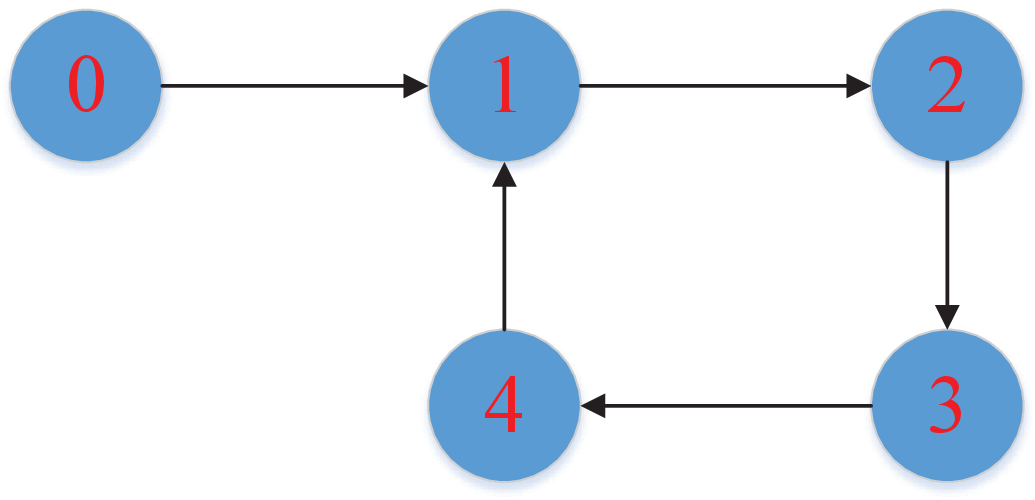}
\caption{Topology of the communication graph.}
\end{center}
\end{figure}

\begin{table}[!htbp]%{Table 1}
\centering
\caption{The number of triggering events}
\label{tab:aStrangeTable}
\label{tab:1}
\begin{tabular}{cccc}
\hline\noalign{\smallskip}
\multicolumn{2}{l}{\textbf{\emph{\qquad \qquad $x_1(0.01)$ \, $x_2(0.01)$ \, $x_1(0.1)$ \, $x_2(0.1)$}}}\\
\noalign{\smallskip}\hline\noalign{\smallskip}
Agent$\:$1 & $\:335$ \qquad $\:440$ \qquad $\:128$  \qquad $\:397$ \\
\noalign{\smallskip}
Agent$\:$2 & $\:324$ \qquad $\:424$ \qquad $\:138$ \qquad $\:391$\\
\noalign{\smallskip}
Agent$\:$3 & $\:305$ \qquad $\:359$ \qquad $\:117$ \qquad $\:342$\\
\noalign{\smallskip}
Agent$\:$4 & $\:260$ \qquad $\:288$ \qquad $\:106$ \qquad $\:309$\\
\noalign{\smallskip}
Leader$\:$0& $\:175$ \qquad $\:175$ \qquad $\:175$ \qquad $\:175$\\
\noalign{\smallskip}\hline
\end{tabular}
\end{table}

\begin{figure*}[t]
\centering{
\subfigure[The trajectories of output $x_{i,1}$.]
{\includegraphics[width=2.35in]{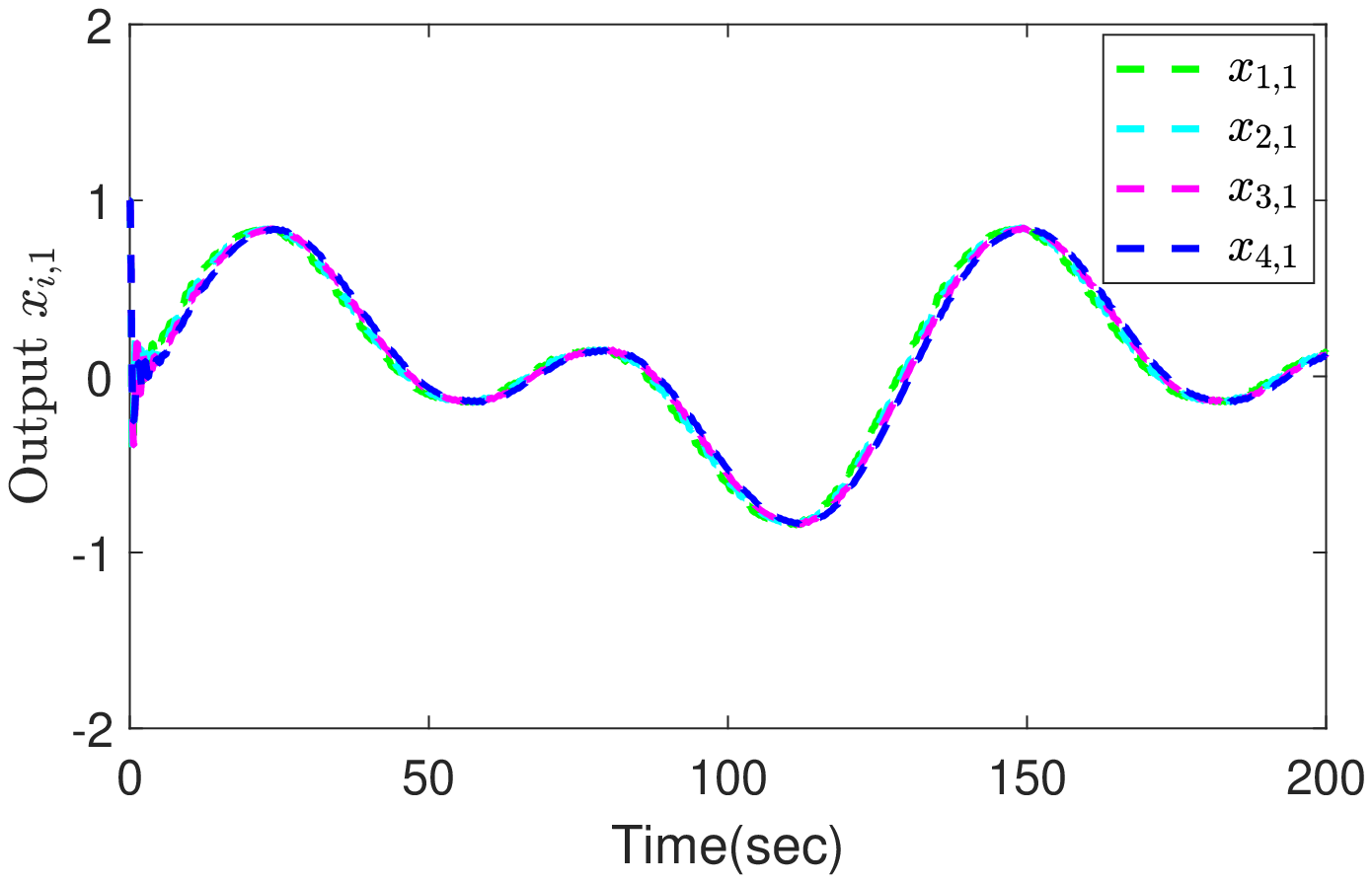}}
\subfigure[Output tracking error $\varepsilon_i$.]
{\includegraphics[width=2.35in]{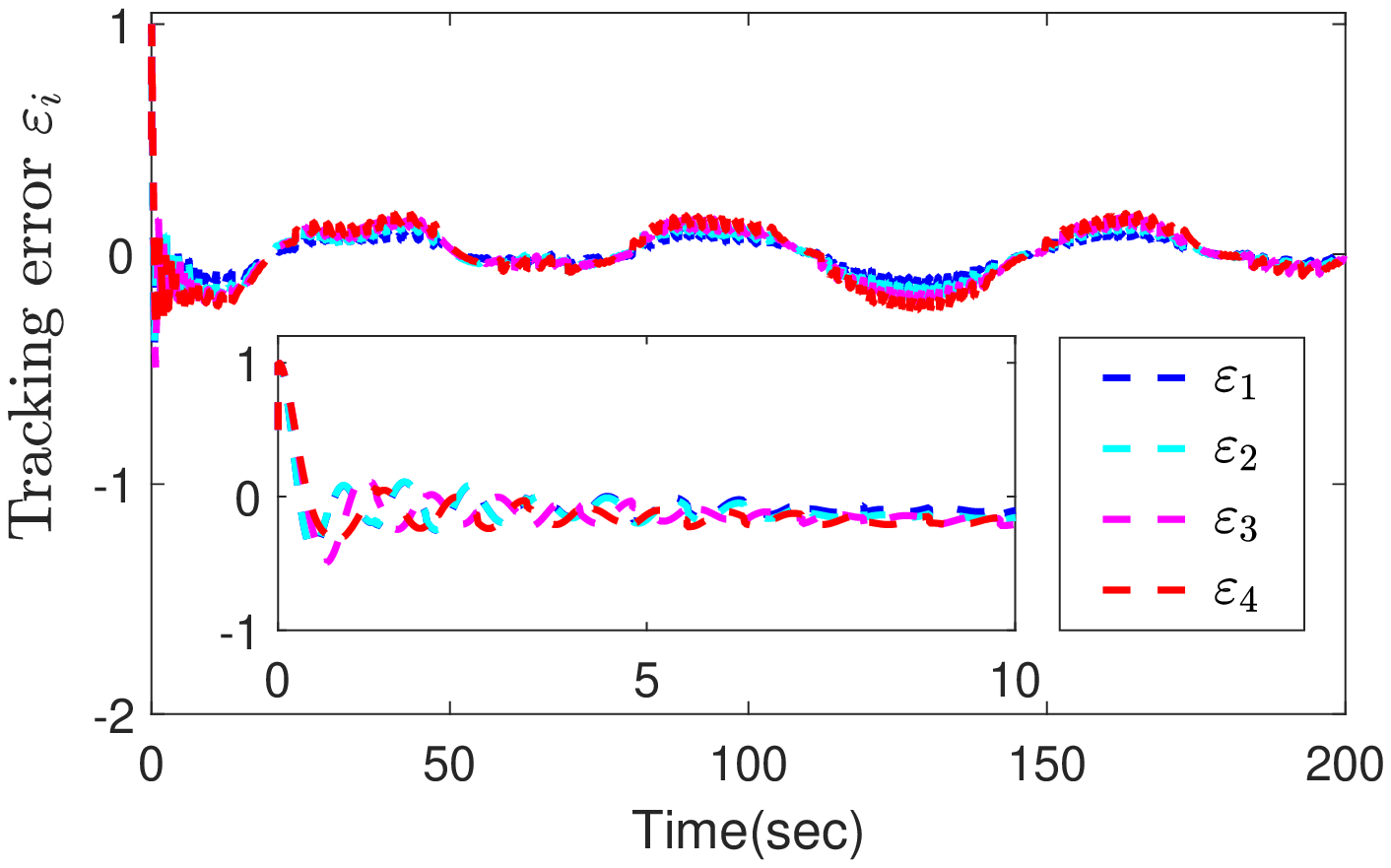}}
\subfigure[Control input $u_i$.]
{\includegraphics[width=2.35in]{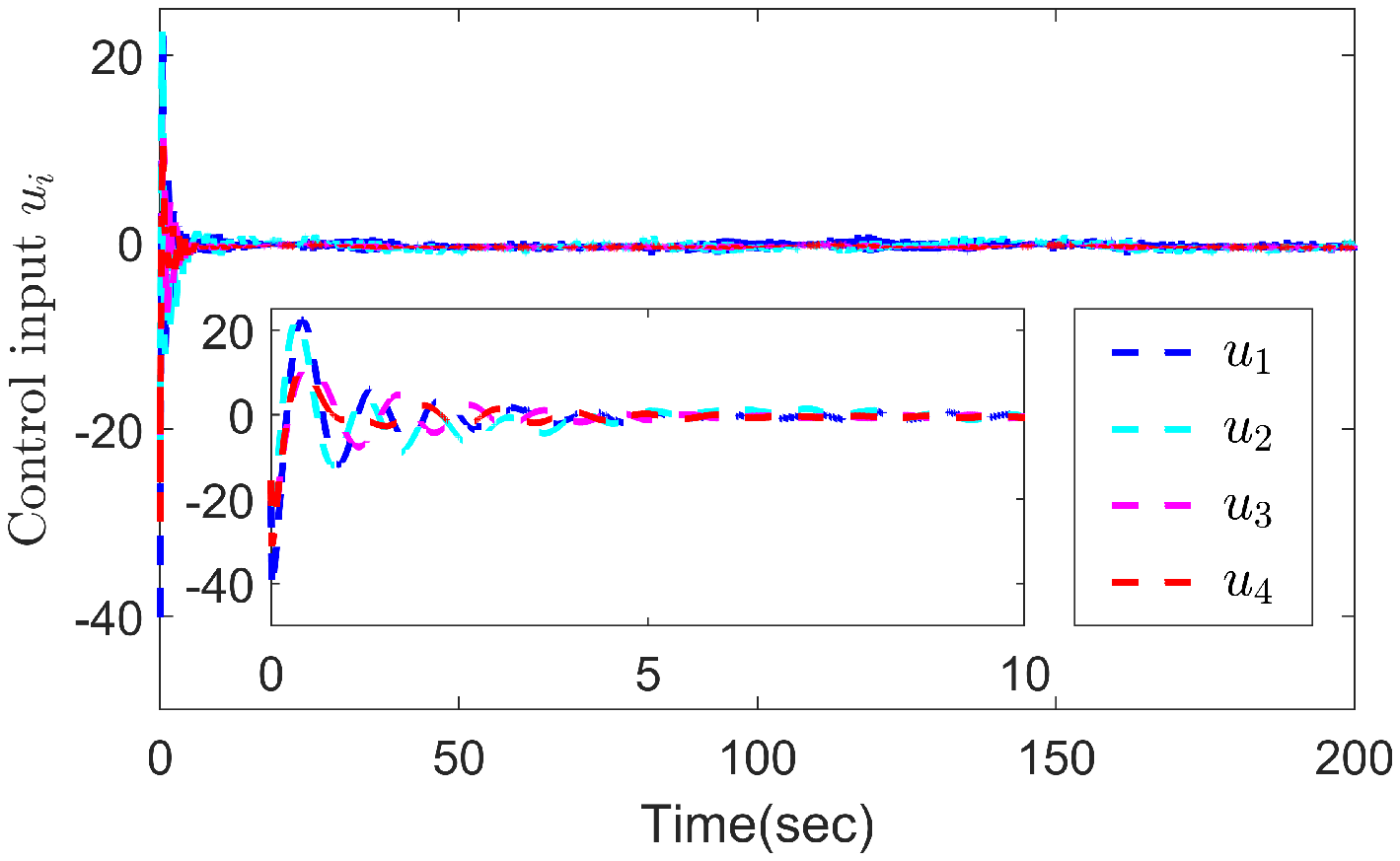}}
\subfigure[Triggering times of $x_{i,1}$ and $x_{i,2}$.]
{\includegraphics[width=2.35in]{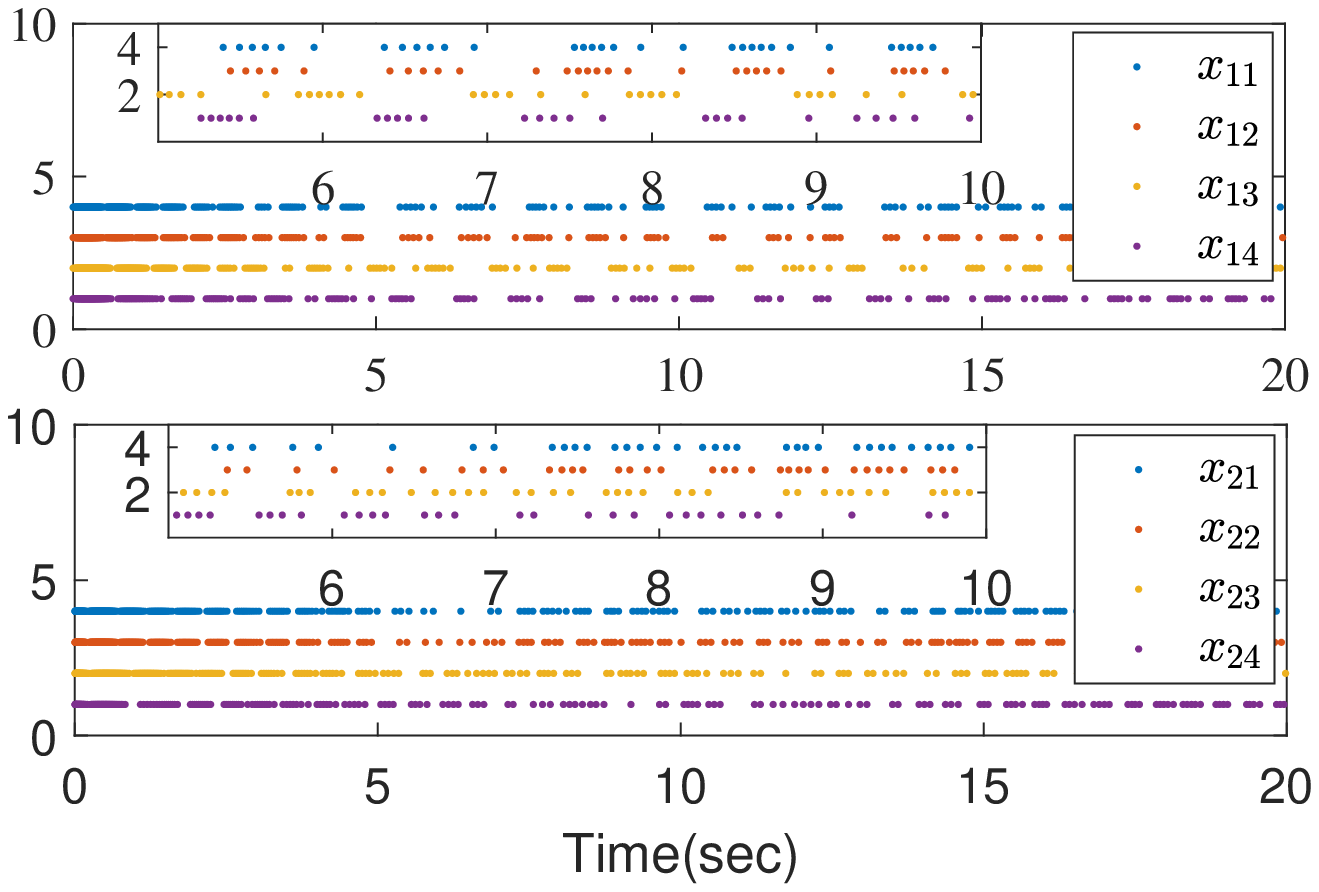}}
\subfigure[Output tracking error $\varepsilon_i$ for the case of increasing triggering thresholds.]
{\includegraphics[width=2.35in]{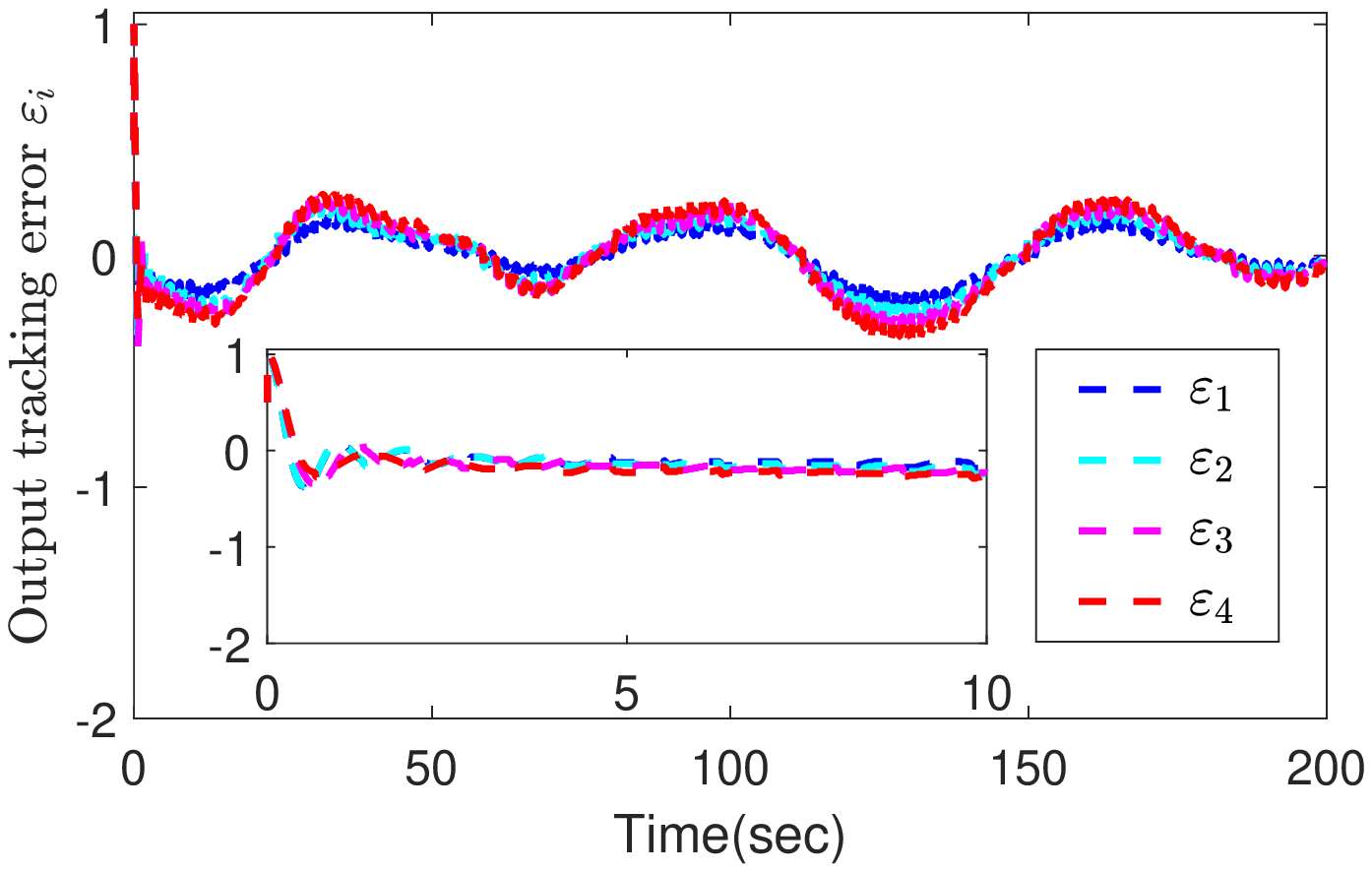}}
\subfigure[Triggering times of $x_{i,1}$ and $x_{i,2}$ for the case of increasing triggering thresholds.]
{\includegraphics[width=2.35in]{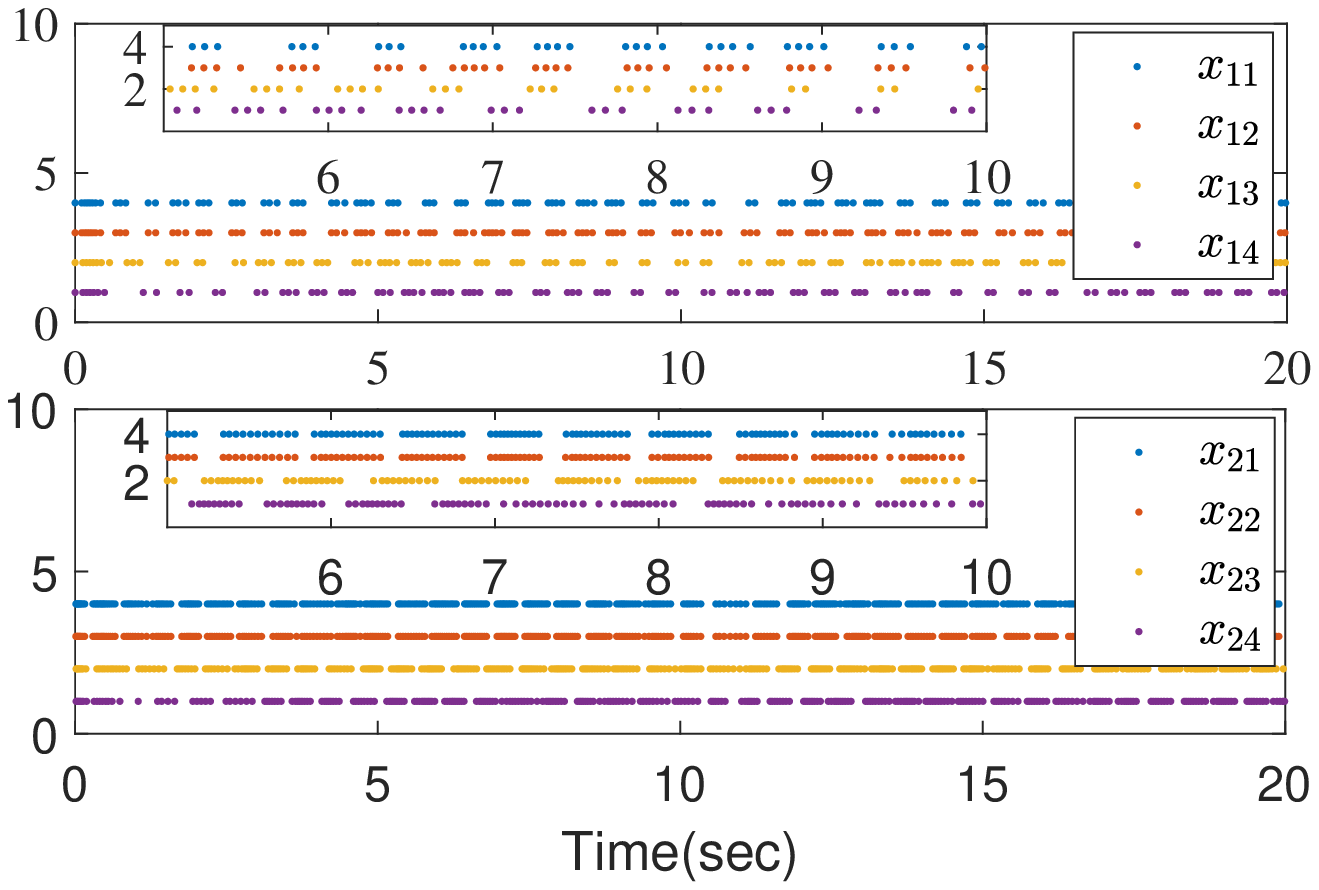}}}
\caption{Simulation results by using the proposed distributed control scheme.}
 \label{simulation_results}
\end{figure*}

\section{Conclusion}%{\color{black}{}}
This work investigates the distributed tracking control problem for a class of networked nonlinear strict-feedback systems with mismatched uncertainties. A fully distributed neuroadaptive control scheme base on intermittent state feedback is derived. The major technical challenge in developing such control strategy is to circumvent the non-differentiability problem of the virtual control arising from state-triggering. By using the results established in the lemmas, it is shown that with the proposed intermittent states only, all the closed-loop signals are SGUUB, with the output tracking error converging to a residual set around zero. Besides, with proper choices of the design parameters, the tracking performance in the mean square sense can be improved. An interesting topic for future research is the consideration of time-varying triggering thresholds with intermittent output feedback only.

\begin{appendices}
\section{}
\textbf{Proof of Theorem 1}.
$\textbf{Step 1}$:
Consider a Lyapunov function =${V_1}= \sum\nolimits_{i = 1}^N\frac{1}{2} z_{i,1}^2 +\sum\nolimits_{i = 1}^N\frac{1}{2} \tilde W_{i,1}^T\Gamma _{i,1}^{ - 1}{{\tilde W}_{i,1}} +\sum\nolimits_{i = 1}^N \frac{1}{2}y_{i,2}^2 + \sum\nolimits_{i = 1}^N \frac{{{\kappa _1}}}{{{2\gamma _{yi0}}}}\tilde y_{i,0}^2$.
Since ${\dot z_{i,1}} = {z_{i,2}} + {\alpha _{i,1}} + {y_{i,2}} + {F_{i,1}}\left( {{\chi_{i,1}}} \right) - {\dot {\hat y}_{i,0}}$, where ${F_{i,1}}\left( {{\chi _{i,1}}} \right) = {f_{i,1}}\left( {{x_{i,1}}} \right)$ is approximated by using the RBFNN on a compact set $\chi _{i,1}=[x_{i,1}, \hat y_{i,0}]^{T} \in{\Omega_{\chi_{i,1}}}$, then it is seen from  (\ref{eq:21}), (\ref{eq:22}) and (\ref{eq:23}) that
\begin{flalign}
{{\dot V}_1}\le& - \frac{{{\kappa _1}{\lambda _{\min }}\left( Q \right)}}{2}{\left\| \varepsilon  \right\|^2} - \sum\limits_{i = 1}^N {{c_{i,1}}z_{i,1}^2}  + \sum\limits_{i = 1}^N {{z_{i,1}}{z_{i,2}}} & \nonumber\\
&- \sum\limits_{i = 1}^N {\frac{{{\sigma _{i,1}}}}{2}\tilde W_{i,1}^T{{\tilde W}_{i,1}}}- \sum\limits_{i = 1}^N {\mu _{i,2}^*y_{i,2}^2}+ \sum\limits_{i = 1}^N {l_{i,1}^2}& \nonumber\\
&- \sum\limits_{i = 1}^N {\frac{{{\kappa _1}{\sigma _{yi0}}}}{4}\tilde y_{i,0}^2} + {\Delta _1}&\label{eq:28}
\end{flalign}
where ${l_{i,1}} {\rm{=}}  - {\dot \alpha _{i,1}}$, $\mu_{i,2}^*$ is a positive design parameter, with $1/{\mu_{i,2}}\ge \mu_{i,2}^*+ 3/4$, and ${\Delta _1} = \sum\nolimits_{i = 1}^N{\frac{{{\kappa _1}{\sigma _{yi0}}}}{2}y_0^2} + \sum\nolimits_{i = 1}^N{\frac{{{\kappa _1}}}{{\gamma _{yi0}^2{\sigma _{yi0}}}}{Y_0}^2}  + \sum\nolimits_{i = 1}^N{\frac{{{\sigma _{i,1}}}}{2}{{\left\| {{W_{i,1}}} \right\|}^2}}  +\sum\nolimits_{i = 1}^N\frac{1}{2}\varepsilon _{i,m}^2$.

$\textbf{Step \emph{k}}$ $(k=2,\cdots,n-1)$:
Consider a Lyapunov function ${V_k} = {V_{k - 1}} + \sum\nolimits_{i = 1}^N \frac{1}{2}z_{i,k}^2 + \sum\nolimits_{i = 1}^N\frac{1}{2} \tilde W_{i,k}^T\Gamma _{i,k}^{ - 1}{{\tilde W}_{i,k}}+ \sum\nolimits_{i = 1}^N\frac{1}{2} y_{i,k + 1}^2$.
Since ${\dot z_{i,k}}= {z_{i,k + 1}} + {\alpha _{i,k}} + {y_{i,k + 1}} + {F_{i,k}}\left( {{\chi _{i,k}}} \right) - {{\dot \alpha }_{i,kf}}$, where ${F_{i,k}}\left( {{\chi _{i,k}}} \right)= {f_{i,k}}\left( {{x_{i,1}}, \cdots ,{x_{i,k}}} \right)$ is approximated by using the RBFNN on a compact set ${{\chi _{i,k}}}=[{x}_{i,1},\cdots,{x}_{i,k}, \hat {y}_{i,0}]^{T} \in{\Omega_{{\chi _{i,k}}}}$, then from (\ref{eq:45}), (\ref{eq:22}) and (\ref{eq:28}), one has
\begin{flalign}
{{\dot V}_k} \le & - \frac{{{\kappa _1}}}{2}{\lambda _{\min }}\left( Q \right){\left\| \varepsilon  \right\|^2} - \sum\limits_{i = 1}^N {\sum\limits_{r = 1}^k {{c_{i,r}}z_{i,r}^2}  + \sum\limits_{i = 1}^N {{z_{i,k}}{z_{i,k + 1}}} } & \nonumber \\
&- \sum\limits_{i = 1}^N {\sum\limits_{r = 1}^k {\frac{{{\sigma _{i,r}}}}{2}\tilde W_{i,r}^T{{\tilde W}_{i,r}}} }  - \sum\limits_{i = 1}^N {\sum\limits_{r = 1}^k {\mu _{i,r + 1}^*y_{i,r + 1}^2} } & \nonumber \\
&- \sum\limits_{i = 1}^N {\frac{{{\kappa _1}{\sigma _{yi0}}}}{4}\tilde y_{i,0}^2}  + \sum\limits_{i = 1}^N {\sum\limits_{r = 1}^k {l_{i,r}^2} }  + {\Delta _k}& \label{eq:47}
\end{flalign}
where ${l_{i,k}}=- {\dot \alpha _{i,k}}$, $\mu_{i,k+1}^*$ is a positive design parameter, with $1/{\mu_{i,k+1}}\ge \mu_{i,k+1}^*+ 3/4$, and ${\Delta _k} = {\Delta _{k - 1}}+ \sum\nolimits_{i = 1}^N{{\frac{{{\sigma _{i,r}}}}{2}{{\left\| {{W_{i,r}}} \right\|}^2}}+ \sum\nolimits_{i = 1}^N\frac{1}{2}\varepsilon _{i,m}^2}$.

$\textbf{Step \emph{n}}$:
Consider a Lyapunov function ${V_n} = {V_{n - 1}} + \sum\nolimits_{i = 1}^N\frac{1}{2} z_{i,n}^2 + \sum\nolimits_{i = 1}^N\frac{1}{2} \tilde W_{i,n}^T\Gamma _{i,n}^{-1}{{\tilde W}_{i,n}}$. From (\ref{eq:9}), (\ref{eq:53}) and (\ref{eq:22}), one has
\begin{flalign}
{{\dot V}_n} \le& - \frac{{{\kappa _1}}}{2}{\lambda _{\min }}\left( Q \right){\left\| \varepsilon  \right\|^2} - \sum\limits_{i = 1}^N {\sum\limits_{r = 1}^n {{c_{i,r}}z_{i,r}^2}}- \sum\limits_{i = 1}^N {\frac{{{\kappa _1}{\sigma _{yi0}}}}{4}\tilde y_{i,0}^2} & \nonumber\\
& - \sum\limits_{i = 1}^N {\sum\limits_{r = 1}^n {\frac{{{\sigma _{i,r}}}}{2}\tilde W_{i,r}^T{{\tilde W}_{i,r}}}} - \sum\limits_{i = 1}^N {\sum\limits_{r = 1}^{n - 1} {\mu _{i,r + 1}^*y_{i,r + 1}^2} }+{\Delta _n}& \nonumber\\
\le &- \frac{{{\kappa _1}}}{2}{\lambda _{\min }}\left( Q \right){\left\| \varepsilon  \right\|^2}-\lambda V_n+\Delta_n
& \label{eq:55}
\end{flalign}
where the fact that $\left| {{l_{i,r}}} \right| \le \varpi_{i,r}$, $r=1,\cdots,n-1$ is used, $\lambda  = \min\{2{c_{i,1}},\cdots,2{c_{i,n}}, {{\sigma _{i,1}}}/{{\lambda_{\max}}\left\{{\Gamma_{i,1}^{-1}} \right\}},\cdots, {{\sigma _{i,n}}}/{{\lambda _{\max }}\left\{ {\Gamma _{i,n}^{ - 1}} \right\}}$, $2\mu _{i,2}^*,\cdots,$ $2\mu _{i,n}^*, \frac{1}{2}{{\sigma _{yi0}}{\gamma _{yi0}}} \left. {} \right\} $ and ${\Delta _n}={\Delta _{n - 1}}+\sum\nolimits_{i = 1}^N {\frac{{{\sigma _{i,n}}}}{2}{{\left\| {{W_{i,n}}} \right\|}^2}}  + \sum\nolimits_{i = 1}^N {\sum\nolimits_{r = 1}^{n-1}} {\varpi_{i,r}^2}+ \sum\nolimits_{i = 1}^N\frac{1}{4}\varepsilon _{i,m}^2$. Thus we can conclude that the results in Theorem 1 are ensured by following the similar analysis in the proof of Theorem 2.

\section{}
\textbf{Proof of Lemma 2}.
Firstly, for $k=1$, we prove that $\left\| {{\phi _{i,1}}\left( {{\chi _{i,1}}} \right) - {\phi _{i,1}}\left( {{{\bar \chi }_{i,1}}} \right)} \right\|\le {\Delta {\phi _{i,1}}}$.
As ${\phi _{i,1}}= {\left[{\phi _{i,11}}, \cdots ,{\phi _{i,1p}}\right]^T}\in \mathcal{R}^{p} $, it holds that $\Delta {\phi _{i,1}} = \left\| {{\phi _{i,1}}\left( {{\chi _{i,1}}} \right) - {\phi _{i,1}}\left( {{{\bar \chi }_{i,1}}} \right)} \right\|\le\sqrt {\Delta \phi _{i,11}^2 +  \cdots  + \Delta \phi _{i,1p}^2}$, where $\Delta\phi _{i,1h}={{\left| {{\phi _{i,1h}}\left( {{\chi _{i,1}}} \right) - {\phi _{i,1h}}\left( {{{\bar \chi }_{i,1}}} \right)} \right|}}$, $h=1,\cdots,p$. Since $\phi_{i,1h}\left( {{\chi _{i,1}}} \right)$ satisfies the locally Lipschitz continuity condition, then we have $\left| { {\phi _{i,1h}}\left( {{\chi _{i,1}}} \right)- {\phi _{i,1h}}\left( {{\bar{\chi} _{i,1}}} \right)} \right| \le L_{1h}\left\|{{\chi _{i,1}}-  {\bar{\chi} _{i,1}}} \right\|$, where $L_{1h}$ is the computable Lipschitz constant.
By the definition of ${\phi_{i,1h}} \left( {{\chi _{i,1}}} \right)$, we know that ${\phi_{i,1h}} \left( {{\chi _{i,1}}} \right)=\exp \left[-{\left(\chi_{i,1}-C_h\right)^{T}\left(\chi _{i,1}-C_h\right)}/ {b_{h}^{2}}\right]$, where ${\chi _{i,1}} = {[\chi _{i,11} , \cdots ,{\chi _{i,1\iota}}]^T}\in \mathcal{R}^{\iota}$ and ${C_h} = [{C_{h1}}, \cdots ,{C_{h\iota}}]^T$ $\in \mathcal{R}^{\iota}$. Taking the derivative of ${\phi_{i,1h}}$ yields ${\dot{\phi}_{i,1h}}=- (2{{\left\| {{{\chi _{i,1}}} - {C_h}} \right\|}^2{({\chi _{i,1}} - C_h)}}/b_h^4)\exp\left[{-{{{{\left\| {{\chi_{i,1}} - {C_h}} \right\|}^2}}} /{b_h^2}}\right]$.
By defining $k_{1h}={\dot{\phi}_{i,1h}}$, $P =\chi_{i,1}-C_h\in \mathcal{R}^{\iota}$, it follows that $k_{1h}=- {A}/{B}$, with $A =2\left\|{P}\right\| ^2P/{b_h^4}\in \mathcal{R}^{\iota}$, $B = \exp\left[-{\left\|{P}\right\|^2} /b_h^2\right]\in \mathcal{R}$. Then it can be computed that
${{\dot k}_{1h}}=-\left({\dot{A}B-A\dot{B}}\right)
/{B^2}= \left\|{P}\right\|^2\left( {4\left\|{P}\right\|^4-6b_h^4} \right)\textbf{\emph{I}}/\left({{b_h^8}} \exp\left(-{{\left\|{P}\right\|^2}}/b_h^2\right)\right)$.
Let ${\dot k_{1h}} = \textbf{0}_{\iota}$, one obtains that $\left\|{P}\right\|^2 = 0$ or $\left\|{P}\right\|^4 = 1.5b_h^4$, i.e., $\left\|{P} \right\| = 0$ or $\left\| {{P}} \right\| = \sqrt[4]{1.5}{b_h}$.
Notice that if $\left\|{P}\right\|= 0$, then $\left\|{k}_{1h}\right\| = 0$, and if $\left\|{P}\right\| = \sqrt[4]{1.5}{b_h}$,  $\left\|{k}_{1h}\right\|= 2{{\sqrt[8]{1.5}}}{\exp\left( \sqrt[4]{1.5}\right)}/{b_h}>0$. Then it is readily seen that $L_{1h}= 2{{\sqrt[8]{1.5}}}{\exp\left( \sqrt[4]{1.5}\right)}/{b_h}$.
Furthermore, it can be derived that $\Delta {\phi _{i,1}}\le \sum\nolimits_{h = 1}^p L_{1h}\Delta \chi _{i,1}$, with $\Delta {\chi_{i,1}} = \left\|{{\chi _{i,1}} - {{\bar \chi }_{i,1}}}\right\|$. Similarly, for $k=2,\cdots,n$, it holds that $\Delta {\phi _{i,k}} \le \sum\nolimits_{h = 1}^p L_{kh}\Delta {\chi_{i,k}}$, where $\Delta {\chi _{i,k}} = \left\|{{\chi _{i,k}} - {{\bar \chi }_{i,k}}}\right\|$. Thus we can derive that $\Delta {\phi _{i,k}}>0$ is a constant that depends on triggering thresholds $\Delta y_0$, $\Delta x_k^i$, $\Delta x_k^{j}$ and design parameters $b_h$, $i=1,\cdots,N$, $j \in \mathcal{N}_i$, $k=1,\cdots,n$, $h=1,\cdots,p$. Notice that \textbf{\emph{I}} is the identity matrix with appropriate dimension, and $\textbf{0}_\iota$ denote the $\iota$-vector of all all zeros. The proof is completed.  $\hfill{\blacksquare}$ %Note that \textbf{\emph{I}} is the identity matrix with appropriate dimension, $\textbf{0}_\iota$ denote the $\iota$-vector of all zeros.

\section{}
\textbf{Proof of Lemma 3}.
From (\ref{eq:10}) and (\ref{eq:59}), it is seen  that
\begin{flalign}
{\left| {{e_i} - {\bar e}_i} \right|} \le & \left( {{d_i} + {\mu _i}} \right)\Delta {x_{1}^{i}} + {d_i}\Delta {x_{1}^j} + {\mu _i}\Delta {y_0}\buildrel \Delta \over =\Delta {e_i}& \label{eq:a1}
\end{flalign}
where $\Delta {y_0} = \left| {{y_0} - {{\bar y}_0}} \right|$. As observed from (\ref{eq:8}), (\ref{eq:9}), (\ref{eq:57}) and (\ref{eq:58}), it follows that
\begin{flalign}
\left| {{z_{i,1}} - {{\bar z}_{i,1}}} \right| =& \left| {{x_{i,1}} - {{\bar x}_{i,1}}} \right| \le \Delta {x_1^{i}}\buildrel \Delta \over = \Delta{z_{i,1}}& \label{eq:a2}\\
\left| {{z_{i,k}} - {{\bar z}_{i,k}}} \right| \le& \left| {{x_{i,k}} - {{\bar x}_{i,k}}} \right| + \left| {{\alpha _{i,kf}} - {{\bar \alpha }_{i,kf}}} \right|& \nonumber\\
\le &\Delta {x_k^{i}} + \Delta {\alpha _{kf}^{i}}\buildrel \Delta \over = \Delta{z_{i,k}},\,k=2,\cdots,n. & \label{eq:a3}
\end{flalign}
In accordance with (\ref{eq:21}) and (\ref{eq:61}), it follows that $\left| {{\alpha _{i,1}} - {{\bar \alpha }_{i,1}}} \right| \le {\kappa _1}\Delta {e_i} + \left( {{c_{i,1}} + 1} \right)\Delta {z_{i,1}}
+ \left| {\hat W_{i,1}^T\left( {{\phi _{i,1}}\left( {{{\bar \chi }_{i,1}}} \right) - {\phi _{i,1}}\left( {{\chi _{i,1}}} \right)} \right)} \right|$.
Invoking \emph{Lemma} 2, it is not difficult to get that $\left| {\hat W_{i,1}^T\left( {{\phi _{i,1}}\left( {{{\bar \chi }_{i,1}}} \right) - {\phi _{i,1}}\left( {{\chi_{i,1}}} \right)}\right)}\right| \le \Delta {\phi _{i,1}} \left\| {{{\tilde W}_{i,1}}} \right\| + \Delta {\phi _{i,1}}\left\| {{W_{i,1}}} \right\|$.
Then we can obtain
\begin{flalign}
&\left| {{\alpha _{i,1}} - {{\bar \alpha }_{i,1}}} \right| \le {\rho _{i,11}}\left\| {{{\tilde W}_{i,1}}} \right\| + {\tau _{i,1}} \buildrel \Delta \over = \Delta {\alpha _{i,1}}& \label{eq:a6}
\end{flalign}
where ${\rho _{i,11}} = \Delta {\phi _{i,1}}$ and ${\tau _{i,1}} = {\kappa _1}\Delta {e_i} + \left( {{c_{i,1}} + 1} \right)\Delta {z_{i,1}} + \Delta {\phi _{i,1}}\left\| {{W_{i,1}}} \right\|$ are positive constants. Using (\ref{eq:45}) and (\ref{eq:62}), one has $\left| {{\alpha _{i,2}} -{{\bar \alpha }_{i,2}}} \right|\le\left( {{c_{i,2}} + 1} \right)\Delta{z_{i,2}}+\Delta {z_{i,1}} + \frac{{\Delta {\alpha_{i,1}}+{\Delta {\alpha _{i,2f}}}}}{{{\mu _{i,2}}}}+ \left| {\hat W_{i,2}^T\left( {{\phi _{i,2}}\left( {{{\bar \chi }_{i,2}}} \right) - {\phi _{i,2}}\left( {{\chi _{i,2}}} \right)} \right)} \right|$. Then it can be derived that
\begin{flalign}
&\left| {{\alpha _{i,2}} - {{\bar \alpha }_{i,2}}} \right| \le \sum\limits_{r = 1}^2 {{\rho _{i,2r}}\left\| {{{\tilde W}_{i,r}}} \right\|}  + {\tau _{i,2}} \buildrel \Delta \over =  \Delta {\alpha _{i,2}}&\label{eq:a8}
\end{flalign}
where ${\rho _{i,21}} = \frac{1}{{{\mu _{i,2}}}}{\rho _{i,11}}$, ${\rho _{i,22}} = \Delta {\phi _{i,2}}$ and ${\tau _{i,2}} = \left( {{c_{i,2}} + 1} \right)\Delta {z_{i,2}} + \Delta {z_{i,1}} + \frac{{{\tau _{i,1}}}}{{{\mu _{i,2}}}} + \frac{{\Delta {\alpha _{i,2f}}}}{{{\mu _{i,2}}}} + \Delta {\phi _{i,2}}\left\| {{W_{i,2}}} \right\|$ are positive constants.
Similarly, for $k=3,\cdots,n$, it holds that
\begin{flalign}
\left| {{\alpha _{i,k}}-{{\bar \alpha }_{i,k}}} \right| \le& \sum\limits_{r = 1}^k {{\rho _{i,kr}}\left\| {{{\tilde W}_{i,r}}} \right\|}  + {\tau _{i,k}} \buildrel \Delta \over =  \Delta {\alpha _{i,k}}&\label{eq:a9}
\end{flalign}
where ${\rho _{i,k1}} = \frac{1}{{{\mu _{i,k}}}}{\rho _{i,\left( {k - 1} \right)1}}$, ${\rho _{i,k2}} = \frac{1}{{{\mu _{i,k}}}}{\rho _{i,\left( {k - 1} \right)2}}, \cdots$, ${\rho _{i,k\left( {k - 1} \right)}} = \frac{1}{{{\mu _{i,k}}}}{\rho _{i,\left( {k - 1} \right)\left( {k - 1} \right)}}$, ${\rho _{i,kk}} = \Delta {\phi _{i,k}}$ and ${\tau _{i,k}} = \left( {{c_{i,k}} + 1} \right)\Delta {z_{i,k}} + \Delta {z_{i,k - 1}} + \frac{{{\tau _{i,k - 1}}}}{{{\mu _{i,k}}}} + \frac{{\Delta {\alpha _{i,kf}}}}{{{\mu _{i,k}}}} + \Delta {\phi _{i,k}}\left\| {{W_{i,k}}} \right\|$ are positive constants.
\end{appendices}

\bibliographystyle{IEEEtran}
\bibliography{event_triggered_20220121}
\end{document}